\documentclass[preprint]{revtex4}
\usepackage{graphicx}
\usepackage{bm}

\usepackage{mhchem}
\usepackage{bm}
\usepackage{amsmath}
\usepackage{amsfonts}
\usepackage{amssymb}

\usepackage{breqn}

\usepackage{balance}

\usepackage{times,mathptmx}

\newcommand{\Id}{{\bm 1}}

\usepackage{subfigure}

\newcommand{\al}{\alpha}

\newcommand{\be}{\begin{equation}}
\newcommand{\ee}{\end{equation}}
\newcommand{\bea}{\begin{eqnarray}}
\newcommand{\eea}{\end{eqnarray}}

\begin{document}

\title{Hybrid Lattice Boltzmann/Finite Difference simulations of viscoelastic multicomponent flows in confined geometries}

\author{A. Gupta$^{1}$, M. Sbragaglia$^{1}$, A. Scagliarini$^{1}$ \\
$^{1}$ Department of Physics and  INFN, University of ``Tor Vergata'', Via della Ricerca Scientifica 1, 00133 Rome, Italy}

\begin{abstract}
We propose numerical simulations of viscoelastic fluids based on a hybrid algorithm combining Lattice-Boltzmann models (LBM) and Finite Differences (FD) schemes, the former used to model the macroscopic hydrodynamic equations, and the latter used to model the polymer dynamics. The kinetics of the polymers is introduced using constitutive equations for viscoelastic fluids with finitely extensible non-linear elastic dumbbells with Peterlin's closure (FENE-P). The numerical model is first benchmarked by characterizing the rheological behaviour of dilute homogeneous solutions in various configurations, including steady shear, elongational flows, transient shear and oscillatory flows. As an upgrade of complexity, we study the model in presence of non-ideal multicomponent interfaces, where immiscibility is introduced in the LBM description using the ``Shan-Chen'' model. The problem of a confined viscoelastic (Newtonian) droplet in a Newtonian (viscoelastic) matrix under simple shear is investigated and numerical results are compared with the predictions of various theoretical models. The proposed numerical simulations explore problems where the capabilities of LBM were never quantified before.
\end{abstract}

\maketitle



\section{Introduction}\label{sec:intro}

Lattice Boltzmann methods (LBM) are nowadays recognized as powerful computational tools for the simulation of hydrodynamic phenomena \cite{Benzi92,Succi01,Gladrow00,ChenDoolen98,Aidun10,Zhang11}. Historically, the main successful applications in the context of computational fluid dynamics pertain the weakly compressible Navier-Stokes equations \cite{Benzi92,Succi01,Gladrow00,ChenDoolen98} and models associated with more complex fluids involving phase transition/separation \cite{SC93,SC94}. However, the spectrum of applications and strengths of LBM in simulating new challenging problems keeps on expanding \cite{Zhang11,Biferale11,Sbragaglia12,SegaSbragaglia13,Varnik11,Dunweg08,KaehlerWagner13}. The LBM does not solve directly the hydrodynamic conservation equations, but rather models the streaming and collision (i.e. relaxation towards local equilibria) of particles, thus offering a series of advantages \cite{Benzi92,Succi01,Gladrow00,ChenDoolen98,Aidun10,Zhang11}. In this paper, we apply the LBM to the simulation of multicomponent viscoelastic fluids. Emulsions or polymer melts, which are present in many industrial and everyday life products, are good examples of such fluids, having the relevant constituents a viscoelastic -rather than a Newtonian- nature \cite{Larson}. We will introduce the kinetics of the polymers using constitutive equations for finitely extensible non-linear elastic dumbbells with Peterlin's closure (FENE-P) \cite{Peterlin61,bird87}, in which the dumbbells can only be stretched by a finite amount, the latter effect parametrized with a maximum extensional length squared $L^2$. The model supports a positive first normal stress difference and a zero second normal stress difference in steady shear. It also supports a thinning effect at large shear, which disappears when $L^2 \gg 1$, a limit where we recover the so-called Oldroyd-B model \cite{Oldroyd50}. Both the FENE-P and Oldroyd-B models have been investigated in many details with other methods based on finite differences \cite{vaithianathan2003numerical,Tome}, finite volumes \cite{Oliveira}, diffuse interface models \cite{Yue04,Yue05}, finite elements \cite{Purnode} and spectral element methods \cite{Chauviere}. There have been already various attempts done with LBM in this direction too. Qian \& Deng \cite{Qian} proposed a modification of the equilibrium distribution to account for the elastic effects, whereas in Ispolatov and Grant \cite{Ispolatov} the elastic effects are taken into account within the framework of a Maxwell model. In Giraud {\it et al.} \cite{Giraud1,Giraud2} and in Lallemand {\it et al.} \cite{Lallemand} LBM schemes for solving the Jeffreys model were proposed, with the hydrodynamic behavior of the LBM emerging with memory effects. In a recent paper, Malaspinas {\it et al.} \cite{Malaspinas10} proposed a new approach to simulate linear and non-linear viscoelastic fluids and in particular those described by the Oldroyd-B and FENE-P constitutive equations. The authors studied and benchmarked the model against various problems, including the 3D Taylor-Green vortex decay, the simplified 2D four-rolls mill, and the 2D Poiseuille flow. A similar approach was used by Denniston {\it et al.} \cite{Denniston} and Marenduzzo {\it et al.} \cite{Marenduzzo} for the simulation of flows of liquid crystals. In other works by Onishi {\it et al.} \cite{Onishi1,Onishi2}, the Fokker-Plank counterpart for the Oldroyd-B and FENE-P models was introduced to carry out simulations with the help of the LBM. The numerical results presented explored the problem of droplet deformation under steady shear. A formulation based on the Fokker-Planck equation was also recently studied by Ansumali \& coworkers \cite{Singh}: the approach was benchmarked by determining the bulk rheological properties for both steady and time-dependent shear and extensional flows, from moderate to large Weissenberg numbers. Finally, we also remark that due to the efficiency of LBM solvers, the latter have been used to replace macroscopic flow solvers for describing dilute polymer solutions \cite{Pham}.\\ 
As witnessed by an increasing amount of works (see \cite{Zhang11} and references therein), LBM has been proven to be particularly suitable to the study of multicomponent systems where interfacial dynamics and phase separation are present, since it can capture basic essential features, even with simplified kinetic models. Significant progress has recently been made in this direction, as evidenced by many LBM that have been developed on the basis of different points of view, including the Gunstensen model \cite{Gunstensen,Liu}, the ``Shan-Chen'' model \cite{SC93,SC94,CHEM09}, the free-energy model \cite{YEO}. However, investigations of viscoelastic flows within the framework of non-ideal multicomponent LBM are rare. The work that better fits these requirements is probably the one by Onishi {\it et al.} \cite{Onishi1,Onishi2}, but the problems there presented suffer of scarce exploration of the effects of confinement and structure of the flow \cite{Greco02,Minale04,ShapiraHaber90,Minale08,Minale10,Minale10review}. Here we go a step forward by presenting a comprehensive study related to the characterization of viscoelastic effects for multicomponent LBM in confined geometries. We numerically and theoretically explore the potentiality of a coupled approach, based on LBM and Finite Difference (FD) schemes, the former used to model two immiscible fluids with variable viscosity ratio, and the latter used to model the polymer dynamics. The numerical model is first benchmarked without phase separation, by characterizing the rheological behaviour of dilute homogeneous solutions with FENE-P model in various steady states (shear and elongational) and transient flows. As an upgrade of complexity, we study the model in presence of non-ideal multicomponent interfaces, where immiscibility is introduced in the LBM description using the ``Shan-Chen'' interaction model \cite{SC93,SC94,CHEM09,SbragagliaBelardinelli}. The problem of a confined viscoelastic (Newtonian) droplet in a Newtonian (viscoelastic) matrix under steady shear is investigated and numerical results are compared with the prediction of various theoretical models. 

\section{Computational Model}\label{sec:model}

In this section we report the essential technical details of the numerical scheme used. We refer the interested reader to the reference papers 
\cite{SC93,SC94,vaithianathan2003numerical,CHEM09,Premnath,Porter,Dunweg07,perlekar06}, where all the details can be found. We consider the Navier-Stokes (NS) and FENE-P  equations for a mixture of two components ($A,B$) in the following form:
\begin{eqnarray}\label{EQ}
\partial_t \rho_{\sigma} + {\bm \nabla} \cdot (\rho_{\sigma} {\bm u}) & = & {\bm \nabla} \cdot {\bm D}_{S,\sigma};  
                                               \hspace{.6in}\sigma=A,B  \label{CONT}\\
\rho \left[ \partial_t \bm u + ({\bm u} \cdot {\bm \nabla}) \bm u \right] & = & -{\bm \nabla} p + {\bm \nabla} \cdot {\bm \sigma}_S + \frac{\eta_P}{\tau_P}{\bm \nabla} \cdot {\bm \sigma}_P +  \sum_{\sigma} {\bm g}_{\sigma};  
                                                 \label{NS}\\
\partial_t {\bm {\mathcal C}} + (\bm u \cdot {\bm \nabla}) {\bm {\mathcal C}}
&=& {\bm {\mathcal C}} \cdot ({\bm \nabla} {\bm u}) + 
                {({\bm \nabla} {\bm u})^T} \cdot {\bm {\mathcal C}} - 
                \frac{{\bm \sigma}_P -{\Id}}{\tau_P}.
                                                   \label{FENE}
\end{eqnarray}
Here, $\rho_{\sigma}$ is the density of the $\sigma$-th component ($\rho=\sum_{\sigma} \rho_{\sigma}$ indicates the total density), ${\bm u}$ represents the baricentric velocity of the mixture, and $p_{\sigma}=c_s^2 \rho_{\sigma}$ ($c_s^2=1/3$ is a constant in the model) is the internal ideal pressure of component $\sigma$, with $p=\sum_{\sigma} p_{\sigma}$.  The diffusion current of one component into the other and the viscous stress tensor of the solvent (S) fluid are
\be\label{DIFFUSIONCURRENT}
{\bm D}_{S,\sigma} = \mu \left[\left({\bm \nabla} p_{\sigma}-\frac{\rho_{\sigma}}{\rho}{\bm \nabla} p\right)-\left({\bm g}_{\sigma}-\frac{\rho_{\sigma}}{\rho}\sum_{\sigma}{\bm g}_{\sigma}\right) \right]
\ee
\be\label{STRESS}
{\bm \sigma}_S=\eta_{s} \left( {\bm \nabla} {\bm u}+({\bm \nabla} {\bm u})^{T}-\frac{2}{3} {\Id} ({\bm \nabla} \cdot {\bm u}) \right) +\eta_{b}{\Id} ({\bm \nabla} \cdot {\bm u}).
\ee
The viscosity coefficients are the shear viscosity $\eta_s$ and the bulk viscosity $\eta_b$, while the coefficient $\mu$ is a mobility parameter regulating the intensity of the diffusion.  The term $\sum_{\sigma} {\bm g}_{\sigma}$ in equation (\ref{NS}) refers to all the contributions coming from internal and external forces. As for the internal forces, we will use the ``Shan-Chen'' interaction model~\cite{SC93} for multicomponent mixtures. The force experienced by the particles of the $\sigma$-th species at ${\bm x}$, is due to the particles of the other species at the neighbouring locations
\begin{equation}\label{eq:SCforce}
{\bm g}_{\sigma}({\bm{x}}) =  - {\cal G} \rho_{\sigma}({\bm{x}}) \sum_{\alpha} \sum_{\sigma'\neq \sigma} w_{\alpha} \rho_{\sigma^{\prime}} ({\bm{x}}+\bm{c}_{\alpha}) {\bm c}_{\alpha} \hspace{.2in} \sigma=A,B
\end{equation}
where ${\cal G}$ is a parameter that regulates the interactions between the two components. The sum in equation (\ref{eq:SCforce}) extends over a set of interaction links $\bm{c}_{\alpha}$ coinciding with those of the LBM dynamics (see below). When the coupling strength parameter ${\cal G}$ is sufficiently large, demixing occurs and the model can describe stable interfaces with a surface tension. The effect of the internal forces can be recast into the gradient of the pressure tensor ${\bm P}^{(int)}$~\cite{SbragagliaBelardinelli}, thus modifying the internal pressure of the model, i.e. ${\bm P} = p \, {\Id}+{\bm P}^{(int)}$, with 
\be\label{PT}
{\bm P}^{(int)}({\bm x}) =\frac{1}{2} {\cal G} \rho_{A}({\bm x})\sum_{\alpha} w_{\alpha} \rho_{B}({\bm x}+{\bm c}_{\alpha}) {\bm c}_{\alpha} {\bm c}_{\alpha}+\frac{1}{2} {\cal G} \rho_{B}({\bm x})\sum_{\alpha} w_{\alpha} \rho_{A}({\bm x}+{\bm c}_{\alpha}){\bm c}_{\alpha} {\bm c}_{\alpha}.
\ee
Upon Taylor expanding the expression (\ref{PT}), we get (explicit dependence on ${\bm x}$ is omitted for simplicity)
\be\label{PTexpanded}
{\bm P} = \left(p+c_s^2{\cal G}\rho_A \rho_B + \frac{1}{4}c_s^4{\cal G}\rho_A \Delta \rho_B+ \frac{1}{4}c_s^4{\cal G}\rho_B \Delta \rho_A  \right) {\Id}+\frac{1}{2} c_s^4{\cal G}\rho_A {\bf \nabla}{\bf \nabla}\rho_B+\frac{1}{2} c_s^4{\cal G}\rho_B {\bf \nabla}{\bf \nabla}\rho_A+{\cal O}({\nabla}^4)
\ee
where we recognize a bulk pressure contribution, $P_b=p+c_s^2{\cal G}\rho_A \rho_B$, and other contributions which are proportional to the derivatives of both densities. The gradient terms establish a diffuse interface whenever phase separation is achieved in the model \cite{CHEM09}. Consistently, the term  ${\bm g}_{\sigma}$ in \eqref{NS}-\eqref{DIFFUSIONCURRENT} may be viewed with its associated Taylor expansion 
\be
{\bm g}_{\sigma} =  - c_s^2 {\cal G} \rho_{\sigma} {\bf \nabla} \rho_{\sigma^{\prime}}- \frac{c_s^4}{2} {\cal G} \rho_{\sigma} \Delta {\bf \nabla} \rho_{\sigma^{\prime}}+{\cal O}({\nabla}^5).
\ee
We refer the interest reader to \cite{SbragagliaBelardinelli}, for a detailed discussion on the relation between the force ${\bm g}_{\sigma}$ and the lattice pressure tensor ${\bm P}$. We wish to stress that the equilibrium properties of the model can also be reformulated in the framework of a free energy model \cite{CHEM09,scarbolo}. In particular, with such formulation, the square bracket of equation \eqref{DIFFUSIONCURRENT} would become proportional gradient of the associated chemical potential, thus being compliant with a thermodynamic framework, where the diffusion force is established by inhomogeneities in the chemical potential. More details can be found in \cite{scarbolo}.\\
A proper tuning of the density gradients in contact with the wall allows to model the wetting properties. In all simulations described in this paper, the resulting contact angle for a droplet placed in contact with the solid walls is $\theta_{eq}=90^{\circ}$ (i.e. neutral wetting). \\
As for the polymer details in equations (\ref{NS}) and (\ref{FENE}), ${\bm {\mathcal C}} \equiv \langle {\bm {\mathcal R}} {\bm {\mathcal R}} \rangle$ is the polymer-conformation tensor, i.e., the ensemble average of the tensor product of the end-to-end distance vector ${\bm {\mathcal R}}$, normalized in such a way that ${\bm {\mathcal C}}$ equals the identity  tensor (${\bm {\mathcal C}}=\Id$) at equilibrium, $\eta_P$ is the viscosity parameter for the FENE-P solute and $\tau_P$ the polymer relaxation time. The polymer feedback into the fluid is parametrized by $\frac{\eta_P}{\tau_P}{\bm \sigma}_P=\frac{\eta_P}{\tau_P}f(r_P) {\bm {\mathcal C}}$, being ${\bm \sigma}_P=f(r_P) {\bm {\mathcal C}}$ the dimensionless counterpart. The FENE-P potential is encoded in $f(r_P)\equiv (L^2 -3)/(L^2 - r_P^2)$, which ensures finite extensibility; $ r_P \equiv \sqrt{Tr({\bm {\mathcal C}})}$ and $L$ are the trace and the (dimensionless) maximum possible extension, respectively, of the polymers \cite{bird87}. As $L$ decreases, the polymer dumbbell becomes less extensible and the maximum level of stress attainable is reduced. In a homogeneous steady uniaxial extension, the extensional viscosity of the polymers increases proportionally to the maximum dumbbell length squared and it becomes infinite in the limit $L^2 \gg 1 $ \cite{Oldroyd50} (see subsection (\ref{simpleelongational})).\\ 
The fluid part of the model (equation (\ref{NS})) is obtained from LBM featuring a multiple relaxation time scheme (MRT). Further technical details of the algorithm can be found in \cite{Premnath,Porter,Dunweg07}, here we just report the essential features of the model. The LBM equation considers the probability density function, $f^{(\sigma)}_{\al}({\bm x},t)$, to find a particle of component $\sigma$ in the space-time location $({\bm x},t)$ with discrete velocity ${\bm c}_{\al}$. In a unitary time lapse, the evolution equation for $f^{(\sigma)}_{\al}({\bm x},t)$ is (double indexes are meant summed upon)
\be\label{LB}
f^{(\sigma)}_{\al}({\bm x}+{\bm c}_{\al},t+1)-f^{(\sigma)}_{\al}({\bm x},t)=- \Lambda_{\alpha \beta}  \left(f^{(\sigma)}_{\beta}-E^{(\sigma)}_{\beta}(\rho_{\sigma},{\bm u}) \right) +  \left(I_{\alpha \beta}-\frac{1}{2}\Lambda_{\alpha \beta}\right)  S_{\beta}({\bm u},{\bm g}_{\sigma}).
\ee
The equilibrium functions are chosen to be 
\be
E^{(\sigma)}_{\alpha}(\rho,{\bm u})=w_{\alpha} \rho \left[1+\frac{{\bm c}_{\al} \cdot {\bm u}}{c_s^2}+\frac{{\bm u}{\bm u}:({\bm c}_{\al} {\bm c}_{\al}-c_s^2 {\Id} )}{2 c_s^4} \right]
\ee
where the weights $w_{\alpha}$ for the D3Q19 \cite{Premnath} LBM used are
\be
w_{\alpha}= \left\{ \begin{array}{ll}
\frac{1}{3}  & \al=0 \\
\frac{1}{18} & \al=1-6 \\
\frac{1}{36} & \al=7-18. \\
\end{array} \right.
\ee
The relaxation towards equilibrium is regulated by the matrix $\Lambda_{\alpha \beta}$, the same for both species. The source term $S_{\alpha} ({\bm u},{\bm g}_{\sigma})$ is chosen as
\be
S_{\alpha} ({\bm u},{\bm g}_{\sigma})=w_{\al}\left[\frac{({\bm c}_{\alpha}-{\bm u})}{c_s^2}+\frac{({\bm c}_{\alpha} \cdot {\bm u})}{c_s^4} {\bm c}_{\alpha} \right] \cdot {\bm g}_{\sigma}
\ee
and the macroscopic variables are the hydrodynamic density (one for each specie) and the common fluid velocity
\be
\rho_{\sigma}({\bm x},t) = \sum_{\alpha=0}^{18} f^{(\sigma)}_{\al} ({\bm x},t) \hspace{.2in} \rho \tilde{{\bm u}} ({\bm x},t) = \sum_{\sigma} \sum_{\alpha=0}^{18}{\bm c}_{\al} f^{(\sigma)}_{\al}({\bm x},t).
\ee
We also choose the equilibrium velocity as the velocity of the whole fluid plus half of the total forcing contribution, i.e. the standard way to define the hydrodynamic velocity in the lattice Boltzmann scheme \cite{Gladrow00,CHEM09}
\be\label{equilibrium:vel}
{\bm u} ({\bm x},t) = \tilde{{\bm u}} ({\bm x},t)+\frac{\sum_{\sigma} {\bm g}_{\sigma}}{2 \rho}.
\ee
In order to perform the relaxation process towards equilibrium, in the spirit of the MRT models, we need to construct sets of linearly independent moments from the distribution functions in velocity space. The moments are constructed from the distribution function through a transformation matrix ${\cal T}$ comprising a linearly independent set of vectors, i.e. $\hat{\bm f}^{(\sigma)}={\cal T} {\bm f}^{(\sigma)}$, with the transformation matrix ${\cal T}$ suitably constructed in terms of the velocity links \cite{Premnath,Porter,Dunweg07}. In the moments space, the collisional operator $\Lambda_{\alpha \beta}$ in the lattice Boltzmann equation (\ref{LB}) is diagonal, thus offering the particular advantage to relax the various processes (diffusive processes and viscous processes) independently.  The relaxation times of the momentum ($\tau_M$), bulk ($\tau_b$) and shear ($\tau_s$) modes in (\ref{LB}) are indeed related to the transport coefficients of hydrodynamics as (The relaxation times for the non-hydrodynamic modes are kept fixed to unitary values)
\begin{equation}\label{TRANSPORTCOEFF}
\mu=\left(\tau_M-\frac{1}{2} \right) \hspace{.2in} \eta_s=\rho c_s^2 \left(\tau_s-\frac{1}{2} \right) \hspace{.2in} \eta_b=\frac{2}{3}\rho c_s^2  \left(\tau_b-\frac{1}{2} \right).
\end{equation}
Some of the modes (${\bm \Pi}_{\sigma}^{(eq)}$ (We refer to $e^{eq}$, $e^{2,eq}$, $p^{eq}_{xx}$, $p^{eq}_{ww}$, $p^{eq}_{xy}$, $p^{eq}_{yz}$, $p^{eq}_{xz}$ defined soon after equation (26) for the D3Q19 model in \cite{Premnath}) of the equilibrium distribution functions $E^{(\sigma)}_{\alpha}(\rho_{\sigma},{\bm u})$ are explicitly affected by the second order tensor of the distribution \cite{Premnath,Porter,Dunweg07}. The polymer stress $\frac{\eta_P}{\tau_P}{\bm \sigma}_P=\frac{\eta_P}{\tau_P}f(r_P){\bm {\mathcal C}}$ appearing in equation (\ref{FENE}) is then added to these modes with a weight that depends on the species, i.e. 
\be\label{MODEtransformation}
{\bm \Pi}^{(eq)}_{\sigma}={\bm \Pi}^{(eq)}_{\sigma}-\frac{\rho_{\sigma}}{\rho}\frac{\eta_P}{\tau_P}f(r_P) {\bm {\mathcal C}}.
\ee
The recovery of the hydrodynamic limit described by equations (\ref{CONT}-\ref{NS}) is ensured by the Chapman-Enskog analysis \cite{Gladrow00,Succi01}. Repeating the calculations reported in \cite{Premnath}, a contribution coming from the polymer stress is found to affect the viscous stress of the equations. Such contribution is measured to be rather small in all the numerical simulations done, ensuring that the balance equations (\ref{CONT}-\ref{FENE}) are reproduced in our simulations. In particular, the weight function $\rho_{\sigma}/\rho$ ensures that the global momentum balance equation (\ref{NS}) has the total stress $\frac{\eta_P}{\tau_P}f(r_P) {\bm {\mathcal C}}$ in the rhs. The idea of changing the lattice Boltzmann stress with a contribution directly related to the polymers feedback stress echoes the work by Onishi {\it et al.} \cite{Onishi1,Onishi2}, although the authors there used a simple single relaxation time scheme. A comprehensive comparison with the results of  Onishi {\it et al.} \cite{Onishi2} is discussed in \ref{AppendixB}. We also remark that the very rich survey of numerical simulations explored in this paper revealed that the idea of changing the lattice Boltzmann stress with a polymer contribution is much more stable than applying the polymer feedback stress as a force term in the LBM. The technical reason of this enhanced stability is presently not understood from the analytical point of view, although it is surely motivating for dedicated studies for future publications.\\ 
The relaxation frequencies in (\ref{TRANSPORTCOEFF}) are chosen in such a  way that $\tau_M=1.0$ lbu (lattice Boltzmann units) and $\tau_s=\tau_b$, corresponding to $\frac{2}{3} \eta_s=\eta_b$ in equation (\ref{STRESS}). The viscosity ratio of the Lattice Boltzmann fluid is changed by letting $\tau_s$ depend on space 
\be
\rho c_s^2 \left(\tau_s-\frac{1}{2} \right)=\eta_s=\eta_A f_+(\phi)+\eta_B f_{-}(\phi)
\ee
where $\phi=\phi({\bm x})=\frac {\rho_A({\bm x})-\rho_B({\bm x})}{\rho_A({\bm x})+\rho_B({\bm x})}$ represents the order parameter. We have indicated with $\eta_{A,B}$ the shear viscosities in the regions with a majority of one of the two components ($A$ or $B$). The functions $f_{\pm}(\phi)$ are chosen as
\be\label{smoothingvisco}
f_{\pm}(\phi)=\left(\frac{1 \pm \tanh(\phi/\Delta)}{2} \right).
\ee
The smoothing parameter $\Delta=0.1$ is chosen sufficiently small so as to recover a matching with analytical predictions for droplet deformation and orientation in shear flow (see \ref{AppendixA}).\\
As for the polymer constitutive equation, we are following the two references~\cite{vaithianathan2003numerical,perlekar06} to solve the FENE-P equation (\ref{FENE}). We maintain the symmetric-positive-definite (SPD) nature of conformation tensor at all times by using the Cholesky-decomposition scheme~\cite{vaithianathan2003numerical,perlekar06}. This addresses two difficulties found in earlier formulations. First, the polymer extension, represented by the trace of the conformation tensor, can numerically exceed the finite extensibility length causing the restoring spring force to change sign and the calculation to rapidly diverge. In the Cholesky decomposition scheme, the conformation tensor is redefined so that this possibility no longer exists.  Secondly, the conformation tensor must remain symmetric and positive definite at all times for the calculation to remain stable. Technically speaking, we first consider the equation for ${\bm \sigma}_P=f(r_P){\bm {\mathcal C}}$. Since ${\bm {\mathcal C}}$ and hence ${\bm \sigma}_P$ are SPD matrices, we can write ${\bm \sigma}_P={\bm {\mathcal L}}{\bm {\mathcal L}}^T$ , where ${\bm {\mathcal L}}$ is a lower-triangular matrix with elements $\ell_{ij}=0$ if $j>i$. Thus, the equation for ${\bm \sigma}_P$ yields an equation set that ensures the SPD of ${\bm {\mathcal C}}$ if $\ell_{ii}>0$~\cite{perlekar06}, a condition which we enforce explicitly by considering the evolution of $\ln \ell_{ii}$ instead of $\ell_{ii}$~\cite{vaithianathan2003numerical}. 
Since the equation for the conformation tensor has no diffusion terms (or other dissipative terms), there is the possibility of generation of sharp gradients (shocks). The Cholesky decomposition scheme eliminates the negative eigenvalues, but to smooth out the shocks in ${\bm {\mathcal C}}$, we add an artificial stress-diffusivity~\cite{vaithianathan2003numerical} term to equation~\eqref{FENE}. We have tested our code with explicit second, fourth and sixth order central finite-difference scheme in space and a second-order Adams-Bashforth method for temporal evolution, finding a stable solution. Hence, we used an explicit second-order central-finite-difference scheme in space to solve the FENE-P equation~\eqref{FENE}. As for the boundary condition for the conformation tensor ${\bm {\mathcal C}}$, we use linear extrapolation at the boundaries.\\
Finally, in order to study separately the effects of matrix and droplet viscoelasticity, we follow the methodologies already developed by Yue {\it et al.} \cite{Yue04}, by allowing the feedback in equation (\ref{NS}) to be  modulated in space with the functions $f_{\pm}(\phi)$.
\be
\rho \left[ \partial_t \bm u + ({\bm u} \cdot {\bm \nabla}) \bm u \right]= -{\bm \nabla} {\bm P}+ {\bm \nabla} \left[(\eta_A f_+(\phi)+\eta_B f_{-}(\phi)) ({\bm \nabla} {\bm u}+({\bm \nabla} {\bm u})^{T} ) \right]+\frac{\eta_P}{\tau_P}{\bm \nabla} [f(r_P){\bm {\mathcal C}} f_{\pm}(\phi) ].
\ee
We remark that other possibilities already exist for implementing the polymer dynamics in LBM \cite{Malaspinas10,Onishi1,Onishi2,Singh}, either by considering directly the evolution equation (\ref{FENE}) \cite{Malaspinas10}, or considering the  the Fokker-Plank counterpart \cite{Onishi1,Onishi2,Singh}. Our algorithm is surely curing problems related to the polymer extension and conformation tensor, which have to remain bounded and positive definite at all times, respectively, for the calculation to remain stable. Nevertheless, we stress that it is not the aim of this paper to propose a comparative study with respect to other existing LBM (or closely related) approaches, as we are interested in assessing the robustness of the methodology in simulating confined problems with multicomponent phases and viscoelastic nature.\\

\section{Homogeneous Dilute Suspensions: Rheology}\label{sec:dilutesuspensions}

In order to validate the numerical scheme described in section \ref{sec:model}, we examined the bulk rheological properties in some canonical steady flow situations, i.e. simple shear flow (section \ref{simpleshear}) and extensional flow (section \ref{simpleelongational}), and also benchmarked time-dependent situations, by verifying the linear viscoelastic behaviour in a small-amplitude oscillatory shearing (section \ref{Gprime}) and the stress relaxation after cessation of a shear flow (section \ref{cessation}) \cite{birdpaper,Herrchen97}. To do that, we switch to zero the coupling constant ${\cal G}$ in equation (\ref{eq:SCforce}), thereby reducing to the case of two miscible gases with an ideal equation of state. We will work with load conditions ensuring very weak compressibility of the system. To properly establish a link between the evolution equation of the conformation tensor (\ref{FENE}) and known results published in the literature \cite{birdpaper,Herrchen97}, we prefer to rewrite the equation for the polymer feedback stress. Starting from the dimensionless polymer feedback stress
\be\label{eq:fluidstress}
{\bm \sigma}_P=f(r_P){\bm {\mathcal C}}=\frac{(L^2 -3)}{(L^2 - Tr({\bm {\mathcal C}}))}{\bm {\mathcal C}}
\ee
and taking the trace of equation (\ref{eq:fluidstress}), we find  $Tr({\bm {\mathcal C}})=\frac{L^2 \, Tr({\bm \sigma}_P)}{L^2-3+Tr({\bm \sigma}_P)}$ and the feedback (\ref{eq:fluidstress}) can be rewritten as
\be
{\bm \sigma}_P = \frac{(L^2 -3)}{(L^2 - \frac{L^2Tr({\bm \sigma}_P)}{L^2-3+Tr({\bm \sigma}_P)})}{\bm {\mathcal C}}=\frac{L^2-3+Tr({\bm \sigma}_P)}{L^2}{\mathcal C}=Z(Tr({\bm \sigma}_P)){\bm {\mathcal C}}
\ee
where we have defined $Z(Tr({\bm \sigma}_P))=\frac{L^2-3+Tr({\bm \sigma}_P)}{L^2}$. The equation of the conformation tensor (\ref{FENE}), with the substitution ${\bm {\mathcal C}}={\bm \sigma}_P/Z$, becomes
\begin{equation}\label{EQ4}
\tau_P\left[\frac{1}{Z}D_t {\bm \sigma}_P -\frac{1}{Z}{\bm \sigma}_P \cdot ({\bm \nabla} {\bm u}) - \frac{1}{Z} ({\bm \nabla} {\bm u})^T\cdot {\bm \sigma}_P -\frac{{\bm \sigma}_P}{Z^2}D_t Z \right]= -{\bm \sigma}_P + {\Id}
\end{equation}
or equivalently
\begin{equation}\label{FENErevised}
Z\left({\bm \sigma}_P - {\Id}\right)+\tau_P\left[D_t {\bm \sigma}_P -{\bm \sigma}_P \cdot ({\bm \nabla} {\bm u}) - ({\bm \nabla} {\bm u})^T\cdot {\bm \sigma}_P - {\bm \sigma}_P D_t \log Z \right]=0
\end{equation}
which directly maps into the equation considered by Bird {\it et al.} \cite{birdpaper} (their equation (10) and subsequent developments). In the following sections we provide benchmark tests for various situations. All the analytical results used can be found in other papers \cite{bird87,Lindner03,birdpaper,Herrchen97} and we limit ourself to a brief review for the sake of completeness.

\subsection{Steady Shear Flow}\label{simpleshear}

\begin{figure}[tbp]
\includegraphics[scale=0.7]{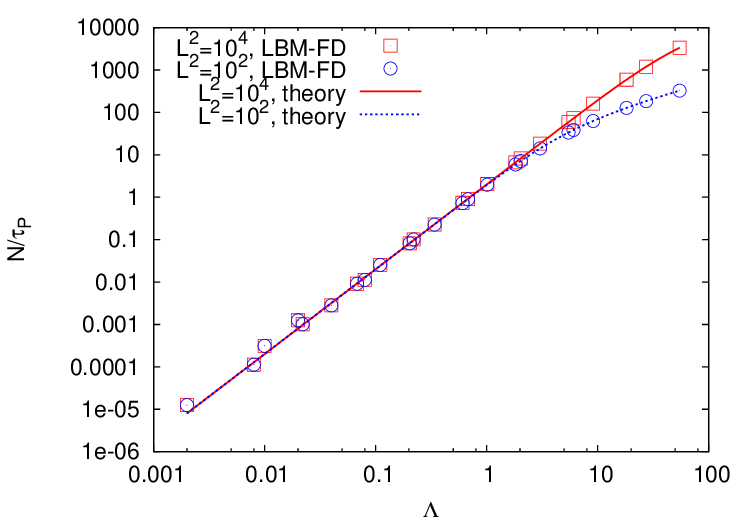}
\includegraphics[scale=0.7]{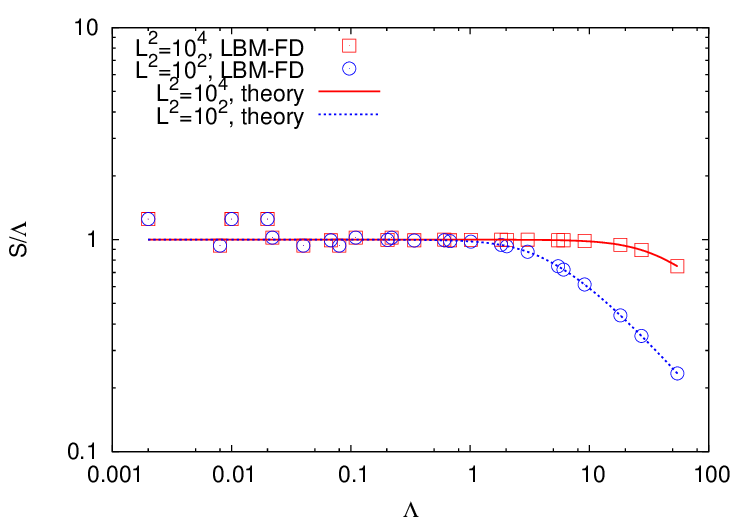}
\caption{We plot the first normal stress difference and the polymer shear viscosity (both scaled with the viscosity $\eta_P$) as a function of the dimensionless shear $\Lambda=\tau_P \dot{\gamma}$. Symbols are the results of the LBM-FD simulations with different imposed shears, different $\tau_P$ and different $L^2$ (see text for details). All the numerical simulations collapse on different master curves, dependently on the value of $L^2$: $L^2=10^2$ (circles) and $L^2=10^4$ (squares). The lines are the theoretical predictions based on equations (\ref{lindner}) and (\ref{lindner2}).}
\label{fig:shearnormalstress}
\end{figure}

We consider equation (\ref{FENErevised}) under the effect of a homogeneous shear flow, $u_x=\dot{\gamma} y$, $u_y=0$, $u_z=0$. The equations, written out in components, become
\begin{equation}\label{eq:steadyshear}
Z \begin{pmatrix} \sigma_{P,xx}-1 & \sigma_{P,xy} & 0 \\ \sigma_{P, yx} & \sigma_{P, yy}-1 & 0 \\ 0 & 0 & \sigma_{P,zz}-1 \end{pmatrix}-\tau_P \left[\dot{\gamma} \begin{pmatrix}  2 \sigma_{P, yx} & \sigma_{P, yy} & 0\\ \sigma_{P,yy} & 0 & 0 \\ 0 & 0 & 0\end{pmatrix}\right]=0.
\end{equation}
We find $\sigma_{P, yy}=\sigma_{P, zz}=1$ so that  $Z=\frac{L^2-1+ \sigma_{P, xx}}{L^2}$. The $xx$ and $xy$ components of equation (\ref{eq:steadyshear}) reduce to the system
\be\label{EQNS}
\begin{cases}
\left(1+\frac{N}{L^2}\right)N=2 \Lambda S\\
\left(1+\frac{N}{L^2}\right)S=\Lambda
\end{cases}
\ee
where $N=(\sigma_{P,xx}-1)$, $\Lambda=\tau_P \dot{\gamma}$, $S=\sigma_{P, xy}$. The quantities $N$ and $S$ represent the first {\it normal stress difference} and the polymer shear stress \cite{bird87,birdpaper} developing in steady shear, respectively. The first normal stress difference is a typical signature of viscoelasticity \cite{bird87}, while from the polymer shear stress we can extract (by dividing for the shear rate) the polymer contribution to the shear viscosity. We immediately see from equations (\ref{EQNS}) that the first normal stress difference hinges on the knowledge of the polymer shear stress
\be\label{N2S}
N=2 S^2
\ee
with $S$ satisfying the following equation
\be\label{cubic}
2\frac{S^3}{L^2}+S-\Lambda=0.
\ee
This equation can be solved exactly \cite{bird87,birdpaper,Lindner03} 
\be\label{lindner}
S(\Lambda,L)=2 \left(\frac{L^2}{6} \right)^{1/2} \sinh \left(\frac{1}{3} \mbox{arcsinh} \left(\frac{\Lambda L^2}{4} \left(\frac{L^2}{6}\right)^{-3/2}\right) \right)
\ee
and, from equation (\ref{N2S}) we find $N$ as
\be\label{lindner2}
N(\Lambda,L)=8 \left(\frac{L^2}{6} \right) \sinh^2 \left(\frac{1}{3} \mbox{arcsinh} \left(\frac{\Lambda L^2}{4} \left(\frac{L^2}{6}\right)^{-3/2}\right) \right).
\ee
Going back to equation (\ref{NS}), we see that the polymer shear stress $\frac{\eta_P}{\tau_P} \sigma_{P, xy} = \frac{\eta_P}{\tau_P} S $ produces a constant shear viscosity only in the Oldroyd-B limit ($S \approx \Lambda= \dot{\gamma} \tau_P$ as $L^2 \gg 1$), while thinning effects are present for finite values of $L^2$.\\ 
In figure \ref{fig:shearnormalstress} we present numerical simulations to benchmark these results. The numerical simulations have been carried out in three dimensional domains with $L_x \times H \times L_z=2 \times 60 \times 2$ cells. Periodic conditions are applied in the stream-flow (x) and in the transverse-flow (z) directions. The linear shear flow $u_x=\dot{\gamma} y$, $u_y=u_z=0$ is imposed in the LBM scheme by applying two opposite velocities in the stream-flow direction ($u_x(x,y=0,z)=-u_x(x,y=H,z)=U_w$) at the upper ($y=H$) and lower wall ($y=0$) with the bounce-back rule \cite{Gladrow00}. We next change the shear in the range $10^{-6} \le 2U_w/H \le 10^{-2}$ lbu and the polymer relaxation time in the range $10^3 \le \tau_P \le 10^5$ lbu for two values of the finite extensibility parameter, $L^2=10^2,10^4$, and fixed $\eta_P=0.136$ lbu. In figure \ref{fig:shearnormalstress} we report the first normal stress difference (left panel) and the polymer shear viscosity (right panel), both rescaled with the viscosity $\eta_P$, as a function of the dimensionless shear $\Lambda=\tau_P \dot{\gamma}$. The values of the conformation tensor are taken when the simulation has reached the steady state. All the numerical simulations collapse on different master curves, dependently on the value of $L^2$. The normal stress difference $N$ increases at large $\Lambda$ to exhibit variable levels depending on $L^2$, and consistently with the theoretical prediction of equation (\ref{lindner2}). The dependence of the normal stress $N$ from $L^2$ directly reflects in the presence of thinning effects visible in the plot of the polymer shear viscosity (see right panel of figure \ref{fig:shearnormalstress}). 

\subsection{Steady Elongational Flow}\label{simpleelongational}

\begin{figure}[tbp]
\begin{center}
\includegraphics[scale=0.7]{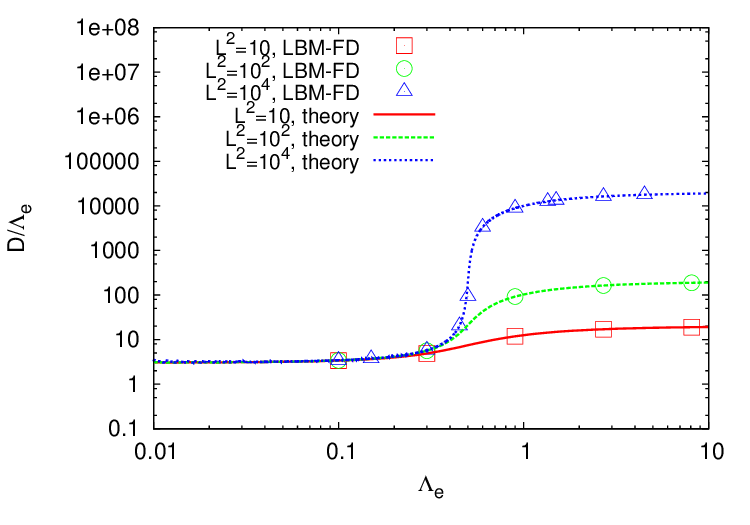}
\caption{We plot the dimensionless elongational viscosity as a function of the dimensionless elongation rate $\Lambda_e=\tau_P \dot{\epsilon}$. Symbols are the results of the LBM-FD numerical simulations with different imposed elongational rates, different $\tau_P$ and different $L^2$ (see text for details).  All the numerical simulations collapse on different master curves, dependently on the value of $L^2$: $L^2=10$ (squares), $L^2=10^2$ (circles) and $L^2=10^4$ (triangles). The lines are the theoretical predictions based on equations (\ref{el8}) and (\ref{el9}).}
\label{fig:elongational}
\end{center}
\end{figure}

We consider equation (\ref{FENErevised}) under the effect of a steady elongational flow, $u_z=\dot{\epsilon} z$, $u_x=-\dot{\epsilon}x/2$, $u_y=-\dot{\epsilon} y/2$, with $\dot{\epsilon}$ the elongation rate. Again, writing out all the components we get
\begin{equation}
Z \begin{pmatrix} \sigma_{P,xx}-1 & 0 & 0 \\ 0 & \sigma_{P,yy}-1 & 0 \\ 0 & 0 & \sigma_{P,zz}-1 \end{pmatrix}+\tau_P \left[ \dot{\epsilon} \begin{pmatrix} \sigma_{P,xx} & 0 & 0\\ 0 & \sigma_{P,yy} & 0 \\ 0 & 0 & -2 \sigma_{P,zz} \end{pmatrix}\right]=0
\end{equation}
implying $\sigma_{P,xx}=\sigma_{P,yy}$. Defining $T=Tr({\bm \sigma}_P)-3$ and $D=\sigma_{P,zz}-\sigma_{P,xx}$, and introducing the dimensionless elongation rate $\Lambda_e=\tau_P\dot{\epsilon}$, we find two independent equations for $D$ and $T$
\be
\begin{cases}
\frac{L^2+T}{L^2}T-2\Lambda_e D  =0\\
-\frac{L^2+T}{L^2} D +\Lambda_e (D+T)+3\Lambda_e  =0
\end{cases}
\ee
which can be rearranged to give us a cubic equation for $D$ as a function of $\Lambda_e$. Such equation is most conveniently written as a quadratic equation in $\Lambda_e$: 
\be\label{el7}
2L^2 D \Lambda_e^2 + \left[-4D^2+(L^2-D-3)(D+3)\right] \Lambda_e + \frac{2D^3}{L^2}-(L^2-D-3)D = 0 
\ee
with associated solutions
\be\label{el8}
(\Lambda_e)_{+,-} = \frac{-P_2 \pm \sqrt{P_2^2-4P_1P_3}}{2P_1}
\ee
where
\be\label{el9}
\begin{cases}
P_1=2DL^2\\
P_2=-4D^2+(L^2-D-3)(D+3)\\
P_3=\frac{2D^3}{L^2}-(L^2-D-3)D.
\end{cases}
\ee
The {\it elongational} viscosity 
\be\label{extensional}
\eta_e=\frac{\eta_P}{\tau_P}\frac{D}{\dot{\epsilon}}
\ee
can be computed by numerically inverting equations (\ref{el8}-\ref{el9}) and paying attention to a proper selection of the sign in equation (\ref{el8}). For small $D$ the solution is given by $(\Lambda_e)_{+}$, as $(\Lambda_e)_{-}$ is negative and divergent. The asymptotic expansion for small $D$ is indeed given by
\be
(\Lambda_e)_{+}= \frac{-P_2 + \sqrt{P_2^2-4P_1P_3}}{2P_1} \approx \frac{D}{3}+{\cal O}(D^2)
\ee
showing that the elongational viscosity approaches a constant value at low elongation rates, which is three times the corresponding zero-shear-rate viscosity. However the radicand of equation (\ref{el8}) is zero when $D=L^2-3$. In such a point, in order to preserve the continuity of the derivative of $\Lambda_e$, we need to consider $(\Lambda_e)_{-}$ as a solution. Consistently, for large $D$, we find
\be
(\Lambda_e)_{-}= \frac{-P_2 - \sqrt{P_2^2-4P_1P_3}}{2P_1} \approx \frac{D}{2L^2}+{\cal O}\left(\frac{1}{D}\right).
\ee
We therefore find the following asymptotic expansion for the elongational viscosity
\be\label{dimensionalelong}
\frac{\eta_e}{\eta_P}= \frac{1}{\tau_P}\frac{D}{\dot{\epsilon}} = \begin{cases} \begin{array}{ll} 3 & \dot{\epsilon} \ll 1 \\ 2 L^2  & \dot{\epsilon} \gg 1 \end{array} \end{cases}
\ee
witnessing a divergence of the elongational viscosity in the Oldroyd-B limit ($L^2 \gg 1$). In figure \ref{fig:elongational} we present numerical simulations to benchmark these results. The numerical simulations have been carried out in a three dimensional cubic domain with edge $H$ consisting of $H \times H \times H = 20 \times 20 \times 20$ cells. Periodic conditions are applied in all directions. The elongational rate is changed in the range $10^{-6} \le \dot{\epsilon} \le 10^{-2}$ lbu and the polymer relaxation time in the range $10^3 \le \tau_P \le 10^5$ lbu, for three values of the finite extensibility parameter, $L^2=10, 10^2,10^4$, and fixed $\eta_P=0.0$ lbu. Again, the values of the conformation tensor are taken when the simulation has reached a steady state. When reporting the quantity $D/\Lambda_e$, i.e. the elongational viscosity scaled by the polymer viscosity, as a function of the dimensionless elongational rate $\Lambda_e$, all the numerical simulations collapse on different master curves, dependently on the value of $L^2$. This behaviour is consistent with the theoretical predictions obtained from equations (\ref{el8}) and (\ref{el9}). For small $\Lambda_e$ the elongational viscosity is just three times the polymer viscosity, while at large $\Lambda_e$ we approach another constant value dependent on the finite extensibility parameter $L^2$ (see equation (\ref{dimensionalelong})).

\subsection{Small amplitude Oscillatory Shearing}\label{Gprime}

\begin{figure}[tbp]
\begin{center}
\includegraphics[scale=0.7]{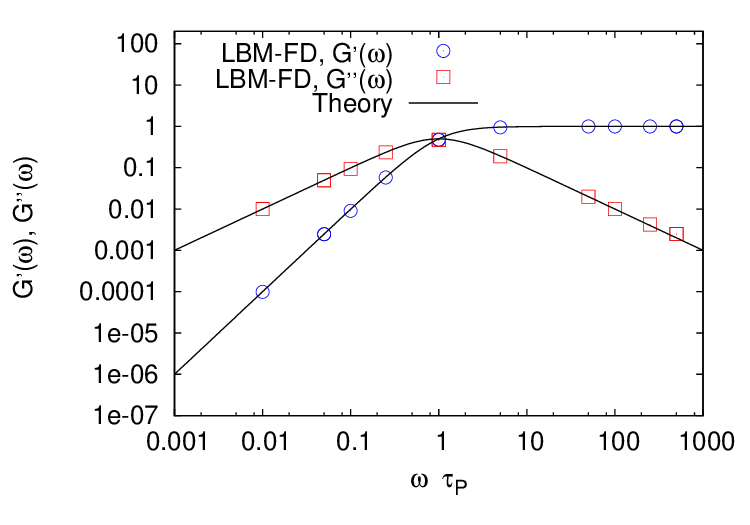}
\caption{We plot the dimensionless storage modulus ($G^{\prime}(\omega)$, circles) and the dimensionless loss modulus ($G^{\prime \prime}(\omega)$, squares) versus the dimensionless frequency $\omega \tau_P$. Results are obtained from the LBM-FD numerical simulations with $L^2=10^5$ (Oldroyd-B limit); black lines show the theoretical prediction for the Oldroyd-B model (see equation (\ref{modulipred})).}
\label{fig:oscillatory}
\end{center}
\end{figure}

By promoting the shear variable considered in section (\ref{simpleshear}) to a time-dependent variable, $u_x=\dot{\gamma}(t) y$, $u_y=0$, $u_z=0$, we can analyze the behaviour of the polymer field under time-dependent loads. We will then analyze the limit of small amplitudes, i.e. $L \gg 1$. In this limit $Z=1$ and we are left with the following time-dependent equation
\begin{equation}
\begin{pmatrix} \sigma_{P,xx}-1 & \sigma_{P,xy} & 0 \\ \sigma_{P,yx} & \sigma_{P,yy}-1 & 0 \\ 0 & 0 & \sigma_{P,zz}-1 \end{pmatrix}+\tau_P \left[\frac{\partial}{\partial t} \begin{pmatrix} \sigma_{P,xx} & \sigma_{P,xy} & 0 \\ \sigma_{P,yx} & \sigma_{P,yy} & 0 \\ 0 & 0 & \sigma_{P,zz} \end{pmatrix} - \dot \gamma(t) \begin{pmatrix}  2 \sigma_{P,yx} & \sigma_{P,yy} & 0\\ \sigma_{P,yy} & 0 & 0 \\ 0 & 0 & 0\end{pmatrix}\right]=0.
\end{equation}
For large $t$, the equations for the first normal stress difference $N$ and polymer shear stress $S$ defined in section (\ref{simpleshear}) are therefore
\be\label{GGG}
\begin{cases}
N+\tau_P\partial_t N =2 \tau_P \dot{\gamma}(t) S\\
S+\tau_P\partial_t S= \tau_P \dot{\gamma}(t).
\end{cases}
\ee
Assuming $\dot{\gamma}(t)=\dot{\gamma}^{(0)} \cos (\omega t)=\Re (\dot{\gamma}^{(0)} e^{-i\omega t})$, we find that the stresses needed to maintain the motion will also be of oscillatory  nature
$$
S=\Re (S^{(0)} e^{-i\omega t})=\Re (\dot{\gamma}^{(0)} \eta^{*} e^{-i\omega t})=\dot{\gamma}^{(0)} \eta^{\prime} \cos (\omega t)- \dot{\gamma}^{(0)} \eta^{\prime \prime} \sin(\omega t) 
$$
where $\eta^{*}=\eta^{\prime}-i\eta^{\prime \prime}$ is the complex viscosity whose components can be computed by taking $S$ and $N$ as complex variables and considering the real and imaginary part of equation (\ref{GGG})
$$
\eta^{\prime}(\omega)=\frac{\tau_P}{1+\omega^2 \tau_P^2}   \hspace{.3in} \eta^{\prime \prime}(\omega)=\frac{\omega \tau^2_P}{1+\omega^2 \tau_P^2}.
$$
The dimensionless storage ($G^{\prime}(\omega)$) and loss ($G^{\prime \prime}(\omega)$) moduli \cite{bird87} are given by
\be\label{modulipred}
G^{\prime \prime}(\omega)=\omega \eta^{\prime}(\omega)=\frac{\tau_P \omega}{1+\omega^2 \tau_P^2}   \hspace{.3in} G^{\prime}(\omega)=\omega \eta^{\prime \prime}(\omega)=\frac{(\omega \tau)^2_P}{1+\omega^2 \tau_P^2}.
\ee
In figure \ref{fig:oscillatory} we present numerical simulations to benchmark these results. The set-up for the numerical simulations is similar to the one presented in section (\ref{simpleshear}), with three dimensional domains consisting of $2 \times H \times 2$ cells, with variable wall-to-wall gap $H$. We then apply an oscillatory shear flow $u_x=\dot{\gamma}(t) y=\frac{2U_w}{H} \cos (\omega t) y$, $u_y=u_z=0$, $\dot{\gamma}(t)=\dot{\gamma}^{(0)} \cos (\omega t)$ at the walls of the LBM simulations and set zero feedback ($\eta_P=0$ lbu) of the polymers into the fluid. The frequency $\omega$ is changed in the range $ 10^{-6} \le \omega \le 10^{-3}$ lbu and the polymer relaxation time in the range $10^3 \le \tau_P \le 10^6$ lbu, for a given value of the finite extensibility parameter, $L^2=10^5$, fixed $\eta_P=0.0$ lbu and maximum wall velocity $U_w=10^{-3}$ lbu. A word of caution is in order, as the assumed flow conditions require that the lattice Boltzmann time to establish a steady shear flow, $\tau_{\nu_S} \sim \frac{H^2}{\nu_S}$ (with $\nu_S$ the solvent kinematic viscosity), is much shorter than the period of the oscillations, i.e.  $\tau_{\nu_S} \omega \ll 1$, otherwise the shear flow will be found in a transient regime. This condition is achieved by a proper tuning of the solvent kinematic viscosity and the wall gap $H$ in all the numerical simulations. As we can see from figure \ref{fig:oscillatory}, the dimensionless storage modulus ($G^\prime(\omega)$) and the dimensionless loss modulus ($G^{\prime\prime}(\omega)$) are in very good agreement with the theoretical prediction of equation (\ref{modulipred}).

\subsection{Stress relaxation after cessation of steady shear flow}\label{cessation}

\begin{figure}[tbp]
\begin{center}
\includegraphics[scale=0.7]{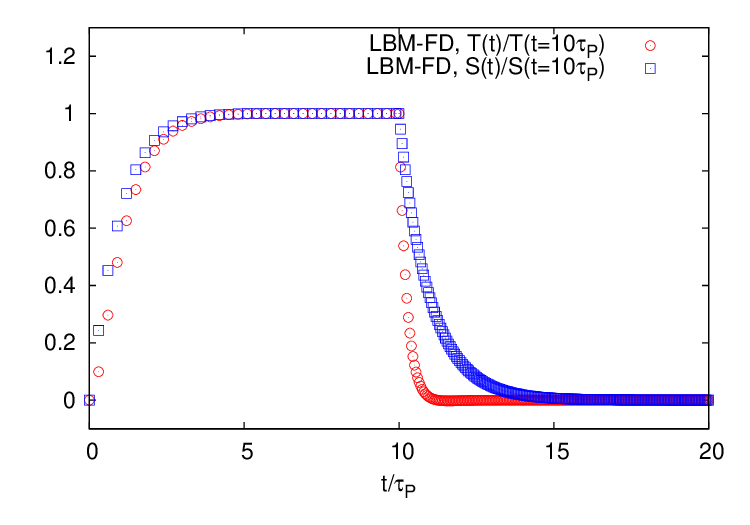}
\includegraphics[scale=0.7]{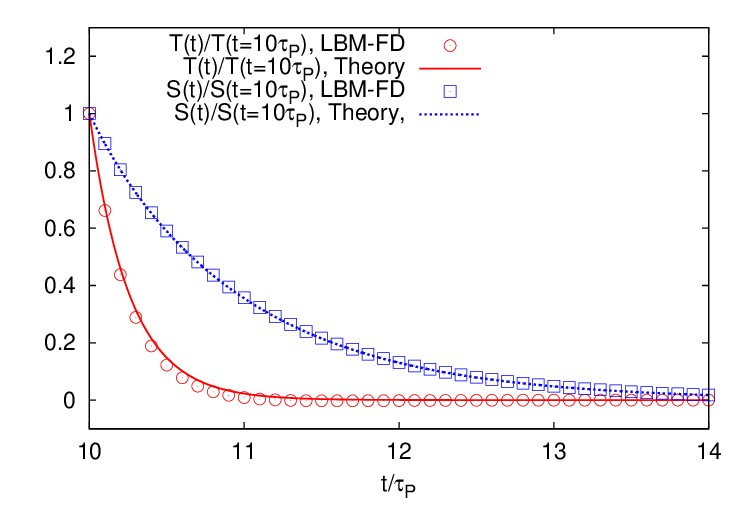}
\caption{We plot the time evolution for the polymer shear stress $S(t)$ (squares) and the excess trace $T(t)=Tr({\bm \sigma}_P)-3$ (circles) versus the dimensionless time ($t/\tau_P$) during the inception of a shear flow and subsequent cessation (see text for details). The shear starts at time $t=0$ and for $t/\tau_P= 10$ the system is under the effect of a steady shear flow. At time  $t/\tau_P=10$ the shear is suddenly switched off and the system starts decaying. The decay process is better illustrated in the right panel where we compare the results of the numerical simulations with the analytical predictions obtained from equations (\ref{timeunclosed}) and (\ref{eq:relax}).}
\label{fig:decay}
\end{center}
\end{figure}

We finally consider a situation with  $u_x=\dot{\gamma}(t) y$, $u_y=0$, $u_z=0$ with $\dot{\gamma}(t)$ being constant for $t<t_0$, and $\dot{\gamma}(t)=0$ for $t \ge t_0$. The equations for $t\ge t_0$ are therefore
\begin{equation}
\begin{aligned}
Z \begin{pmatrix} \sigma_{P,xx}-1 & \sigma_{P,xy} & 0 \\ \sigma_{P,yx} & \sigma_{P,yy}-1 & 0 \\ 0 & 0 & \sigma_{P,zz}-1 \end{pmatrix} + \tau_P \frac{\partial}{\partial t} \begin{pmatrix} \sigma_{P,xx} & \sigma_{P,xy} & 0 \\ \sigma_{P,yx} & \sigma_{P,yy} & 0 \\ 0 & 0 & \sigma_{P,zz} \end{pmatrix} - \tau_P \begin{pmatrix} \sigma_{P,xx} & \sigma_{P,xy} & 0 \\ \sigma_{P,yx} & \sigma_{P,yy} & 0 \\ 0 & 0 & \sigma_{P,zz} \end{pmatrix} D_t \log Z=0.
\end{aligned}
\end{equation}
We next write down the equations for the variables $S=\sigma_{P,xy}$ and $T=Tr({\bm \sigma}_P)-3$ 
\be\label{eq:cess}
\begin{cases}
\frac{L^2+T}{L^2}T+\tau_{P}\partial_t T-\tau_P (3+T) \frac{\partial_t T}{(L^2+T)}=0\\
\frac{L^2+T}{L^2}S+\tau_{P}\partial_t S-\tau_P S  \frac{\partial_t T}{(L^2+T)}=0.
\end{cases}
\ee
The first of equations (\ref{eq:cess}) can be solved to get a differential equation for $T$
\be\label{timeunclosed}
\frac{\partial_{\tilde{t}} T}{T}=\frac{(L^2+T)^2}{L^2(3-L^2)}
\ee
where $\tilde{t}=t/\tau_P$. The Oldroyd-B ($L^2\gg 1$) limit simply implies an exponential decay $T(t)=T_0 e^{-(t-t_0)/\tau_P}$, where with the subscript $0$ we indicate variables at time $\tilde{t}=t_0/\tau_P$. For the general case with finite extensibility parameter $L^2$ in equation (\ref{timeunclosed}), $T(t)$ cannot be written in terms of elementary functions. However, by a proper manipulations of equations (\ref{eq:cess}), it is always possible to get an equation relating the shear stress to the trace of the stress during relaxation \cite{bird87}
\be\label{eq:relax}
\frac{S(t)}{S_0}=\left(\frac{T(t)}{T_0}\right)^{(L^2-3)/L^2}\left(\frac{L^2+T(t)}{L^2+T_0}\right)^{1-(L^2-3)/L^2}.
\ee
For completeness, we note that further manipulations \cite{birdpaper,Herrchen97} of equations (\ref{eq:cess}) allow  to show that the area under the stress-relaxation curve is closely related to the first normal stress-difference before the cessation of the shear flow
\be\label{eq:graal}
N_0 (t<t_0) = 2 \dot{\gamma}\int_{t_0}^{\infty} S dt = 2 \dot{\gamma} \tau_P \int_{t_0/\tau_P}^{\infty} S d \tilde{t}.
\ee
In the left panel of figure \ref{fig:decay} we plot the time evolution for both $S(t)$ and $T(t)$ versus the dimensionless time ($t/\tau_P$) in the process of an inception of shear flow with the approaching to the steady state and subsequent cessation. The set-up for the numerical simulations is similar to the one presented in section (\ref{simpleshear}), with three dimensional domains consisting of $L_x \times H \times L_z=2 \times 60 \times 2$ cells. The shear is set to $2 U_w/H=10^{-3}$ lbu at time $t/\tau_P=0$, with the polymer relaxation time $\tau_P=10^4$ lbu and finite extensibility parameter $L^2=4.1$. The value of $L^2$ is chosen to create a net distinction between the time evolution of $S(t)$ and $T(t)$, that otherwise would be identical in the Oldroyd-B limit ($L^2 \gg 1$, see also equation (\ref{eq:relax})). The feedback of the polymer into the fluid is set to zero. For $t/\tau_P = 10$ (that means $t_0=10 \tau_{P}$ in the above equations) the system is surely under the effect of a steady shear flow. At that time, the shear is suddenly switched off and the system starts decaying. The decay process is illustrated in the right panel of figure \ref{fig:decay}, where we compare the results of the numerical simulations with the analytical predictions obtained from equations (\ref{timeunclosed}) and (\ref{eq:relax}). 

\section{Binary Mixtures with Viscoelastic Phases}\label{sec:binary_on}

In this section we describe problems where both phase segregation and viscoelasticity are present. First of all we switch on immiscibility: when ${\cal G}>{\cal G}_c$ in equation (\ref{eq:SCforce}), with ${\cal G}_c$ a critical value of the coupling constant, the binary mixture separates into two phases, each with a majority of one of the two components and with the interface between the two phases described as a thin layer of thickness $\xi$ across which the fluid properties change smoothly. The values of the interface thickness and the mobility $\mu$ (see equation (\ref{DIFFUSIONCURRENT})) need to be larger than those suggested by physical considerations in order to make the simulations affordable. They are empirically tuned in order to match the analytical predictions of sharp-interface hydrodynamics (see later).\\
We will then apply our numerical approach to the characterization of deformation of droplets in confined geometries, where the involved phases may possess a viscoelastic nature. This is a relevant problem, for example, when determining the properties of emulsions microstructures \cite{Christopher07,Seeman12}. Emulsions play an important role in a huge variety of applications, including foods, cosmetics, chemical and material processing \cite{Larson}. Deformation, break-up and coalescence of droplets occur during flow, and the control over these processes is imperative to synthesize the desired macroscopic behaviour of the emulsion. Most of the times, the synthesis of the emulsion takes place in presence of confinement, and relevant constituents have commonly a viscoelastic -rather than Newtonian- nature. The ``single'' drop problem has been considered to be the simplest model: in the case of dilute emulsions with negligible droplets interactions, the dynamics of a single drop indeed provides complete information about the emulsion behaviour. Single drop deformation has been extensively studied and reviewed in the literature for the case of Newtonian \cite{Taylor34,Grace,Stone,Rallison} and also non-Newtonian fluids \cite{Greco02,Minale04,Minale10,Guido11}. 

\subsection{Effects of confinement on droplet deformation}

\begin{figure}[tbp]
\begin{center}
\includegraphics[scale=0.32]{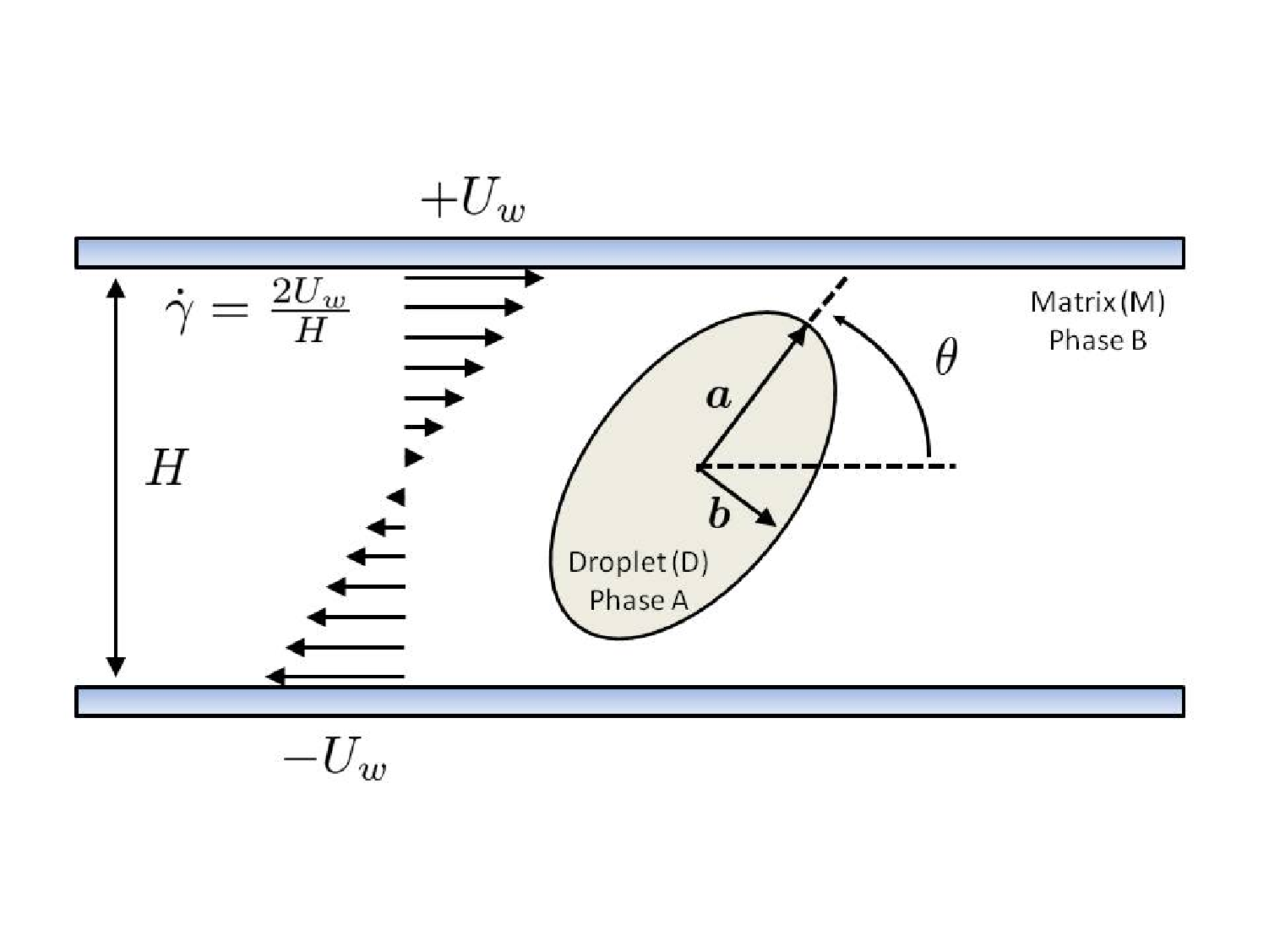}
\includegraphics[scale=0.69]{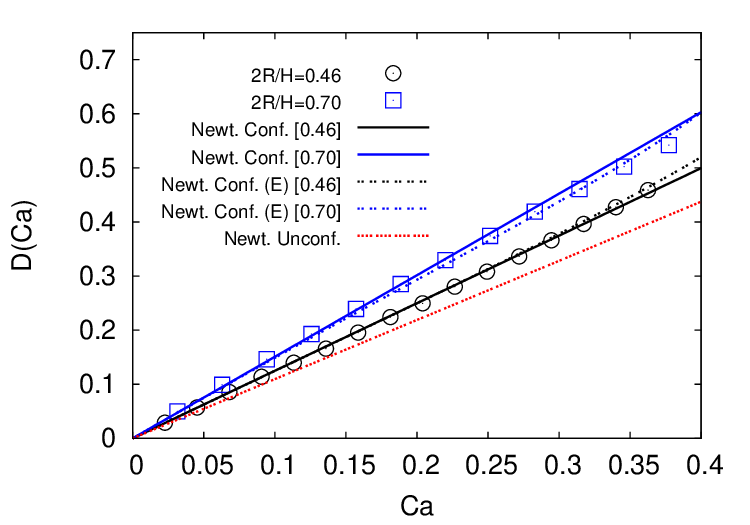}
\caption{Left Panel: shear plane ($z=H/2$) view of the numerical set-up for the study of deformation of confined droplets. A Newtonian droplet (Phase $A$) with radius $R$ and shear viscosity $\eta_A$ is placed in between two parallel plates at distance $H$ in a Newtonian matrix (Phase $B$) with shear viscosity $\eta_B$. We then add a polymer phase with shear viscosity $\eta_P$ in the droplet or matrix phase. We work with unitary viscosity ratio, defined in terms of the total (fluid+polymer) shear viscosity: $\lambda=(\eta_A+\eta_P)/\eta_B=1$, in case of droplet viscoelasticity; $\lambda=\eta_A/(\eta_B+\eta_P)=1$, in case of matrix viscoelasticity. A shear is applied by moving the two plates in opposite directions with velocities $\pm U_w$. Right panel: We report the steady state deformation parameter $D$ for a Newtonian droplet in a Newtonian matrix ($\eta_P=0.0$ lbu) under steady shear as a function of the associated Capillary number $Ca$. For small $Ca$ the linearity of the deformation is captured by the numerical simulations, but the numerical results overestimate Taylor's prediction (referred to as ``Newtonian Unconfined''), being well approximated by the theoretical prediction of Shapira \& Haber for a confined droplet \cite{ShapiraHaber90} (referred to as ``Newtonian confined''). Two confinement ratios are considered: $2R/H=0.46$ and $2R/H=0.7$. We also report the theoretical predictions of the ``ellipsoidal'' models \cite{Minale08,Minale10b} (referred to as ``Newtonian confined (E)''). For the ``confined" theoretical prediction, larger deformations are related to larger confinement ratio. }
\label{fig:deformationNewtonian}
\end{center}
\end{figure}

In the classical problem studied by Taylor \cite{Taylor34}, a droplet with radius $R$, interfacial tension $\sigma_{AB}$, and viscosity $\eta_D$ is suspended in another immiscible fluid matrix with viscosity $\eta_M$ under the effect of a shear flow with intensity $\dot{\gamma}$ (see left panel of figure \ref{fig:deformationNewtonian}). The various physical quantities are grouped in two dimensionless numbers, the Capillary number 
\be\label{capillary}
Ca=\frac{\dot{\gamma} R \eta_M}{\sigma_{AB}}
\ee
giving a dimensionless measure of the balance between viscous and interfacial forces, and the viscosity ratio $\lambda=\eta_D/\eta_M$, going from zero for vanishing values of the droplet viscosity (i.e. a bubble) to infinity in the case of a solid particle.  In order to quantify the deformation of the droplet, we study the deformation parameter $D=(a-b)/(a+b)$, where $a$ and $b$ are the droplet semi-axes in the shear plane, and the orientation angle $\theta$ between the major semi-axis and the flow direction (see also the left panel of figure \ref{fig:deformationNewtonian}). Taylor's result, based on a small deformation perturbation analysis to first-order, relates the deformation parameter to the Capillary number $Ca$, 
\be\label{deformation}
D = \frac{(19\lambda+ 16)}{(16 \lambda+16)} Ca
\ee
whereas the orientation angle is constant and equal to $\theta=\pi/4$ to first order. Taylor's analysis was later extended by working out the perturbation analysis to second order in $Ca$, which leaves unchanged the expression of the deformation parameter (\ref{deformation}) and gives the ${\cal O}(Ca)$ correction to the orientation angle \cite{Rallison2,Chaffey}. The effects of confinement have been theoretically addressed at  ${\cal O}(Ca)$ by Shapira and Haber \cite{ShapiraHaber90,Sibillo06} based on Lorentz's reflection method. They found that the deformation parameter in a confined geometry can be obtained by the Taylor's result through a correction in the power of the ratio between the droplet radius at rest $R$ and gap between the walls $H$
\be
D=\frac{(19\lambda+ 16)}{(16 \lambda+16)}\left[1+C_{sh} \frac{2.5 \lambda+1}{\lambda+1} \left(\frac{R}{H}\right)^3 \right] Ca
\ee
where $C_{sh}$ is a tabulated numerical factor depending on the relative distance between the droplet center and the wall (the value of $C_{sh}$ for droplets placed halfway between the plates is $C_{sh}=5.6996$).\\
LBM have already been used to model the droplet deformation problems \cite{Xi99,VanDerSman08,Komrakovaa13,Liuetal}. Three-dimensional numerical simulations of the classical Taylor's problem \cite{Taylor34} have been performed by Xi \& Duncan \cite{Xi99} using the ``Shan-Chen'' model \cite{SC93,SC94}. The single droplet problem was also investigated by Van Der Sman \& Van Der Graaf \cite{VanDerSman08} using a ``free energy'' LBM. LBM modelling of two phase flows is intrinsically a diffuse interface method and involves a finite thickness of the interface between the two liquids and related free-energy model parameters. These model parameters are characterized by two dimensionless numbers: the P\'{e}clet ($Pe$) and Cahn numbers ($Ch$), the Cahn number is the interface thickness normalized with the droplet radius, whereas the P\'{e}clet number $Pe$ is the ratio between the convective time scale and the interface diffusion. A recent comprehensive study by Komrakova {\it et al.} has investigated the influence of $Pe$, $Ch$ and mesh resolution on the accuracy and stability of the numerical simulations. Drops of moderate resolution (radius less than 30 lattice units) require smaller interface thickness, while a thicker interface should be used for highly resolved drops. Those parameters have to be within certain ranges to reproduce the physical behavior \cite{VanDerSman08,Komrakovaa13} of sharp-interface hydrodynamics \cite{casciola13}. Since our aim is to quantify and explore the importance of viscoelasticity in our simulations, we choose the aforementioned parameters in such a way that the Newtonian (sharp-interface) predictions for droplet orientation and deformation  are well reproduced. All the simulations described in the following sections refer to cases with polymer relaxation times ranging in the interval $0 \le \tau_{P} \le 4000$ lbu and finite extensibility $10 \le L^2 \le 10^4$. The numerical simulations have been carried out in three dimensional domains with $L_x \times H \times H = 288 \times 128 \times 128$ lattice cells. The droplet radius $R$ has been changed in the range $30 \le R \le 40 $ lattice cells with fixed $H$ to achieve different confinement ratios $2R/H$. Periodic conditions are applied in the stream-flow (x) and in the transverse-flow (z) directions. The droplet is subjected to a linear shear flow $u_x=\dot{\gamma} y$, $u_y=u_z=0$, with the shear introduced with two opposite velocities in the stream-flow direction ($u_x(x,y=0,z)=-u_x(x,y=H,z)=U_w$) at the upper ($y=H$) and lower wall ($y=0$). For the numerical simulations presented we have used ${\cal G}=1.5$ lbu in (\ref{eq:SCforce}) (the critical point is at ${\cal G}_c=1.0$ for the parameters chosen) and a total average density of $2.1$ lbu, corresponding to a surface tension $\sigma_{AB}=0.09$ lbu and associated bulk densities $\rho_A=2.0$ lbu and $\rho_B=0.1$ lbu in the $A$-rich region. Some numerical studies to test the sensitivity with respect to a change in the resolution and model parameters used are reported in \ref{AppendixA}.\\
In the right panel of figure \ref{fig:deformationNewtonian} we report the steady state deformation parameter $D$ for a Newtonian droplet under steady shear as a function of the associated Capillary number $Ca$ for two different confinement ratios: $2R/H=0.46$ and $2R/H=0.7$.  The viscosity ratio between the droplet phase and the matrix phase is fixed to $\lambda=\eta_D/\eta_M=\eta_A/\eta_B=1$, with the dynamic viscosities equal to $\eta_A=\eta_B=1.74$ lbu. The linearity of the deformation is captured by the numerical simulations up to the largest $Ca$ considered, but the numerical results overestimate Taylor's prediction (referred to as ``Newtonian Unconfined'') being well approximated by the theoretical prediction of Shapira \& Haber for a confined droplet \cite{ShapiraHaber90} (refereed to as ``Newtonian confined''). Confinement promotes larger deformation and wall effects act to stabilize the resulting elongated drop shapes (which would be otherwise unstable in the unbounded case) by confining the drop within closed streamlines \cite{Sibillo06}. For completeness, we also report a comparison with the steady state deformation prediction of a model proposed recently by Minale \cite{Minale08}, describing the dynamics (and steady states) of a droplet under the assumption that it deforms into an ellipsoid. This model belongs to the family of ``ellipsoidal'' models \cite{Minale10b}, which were originally introduced to describe the dynamics of a single Newtonian drop immersed in a matrix subjected to a generic flow field. The steady state predictions of such models for small $Ca$ are constructed in such a way to recover the exact perturbative result, i.e. Taylor's result for an unbounded droplet \cite{MaffettoneMinale98} or the Shapira \& Haber result for a confined droplet \cite{Minale08}. The prediction of these ellipsoidal models is hardly distinguishable from the perturbative results \cite{ShapiraHaber90} in these Newtonian cases, at least for the range of parameters that we have used in the numerical simulations. Nevertheless, these models will be quite useful when discussing the influence of viscoelasticity on droplet deformation and orientation, as will be done in the following sections.

\subsection{Effects of Viscoelasticity on droplet deformation and orientation}

\begin{figure}[t!]
\begin{center}
\includegraphics[scale=0.7]{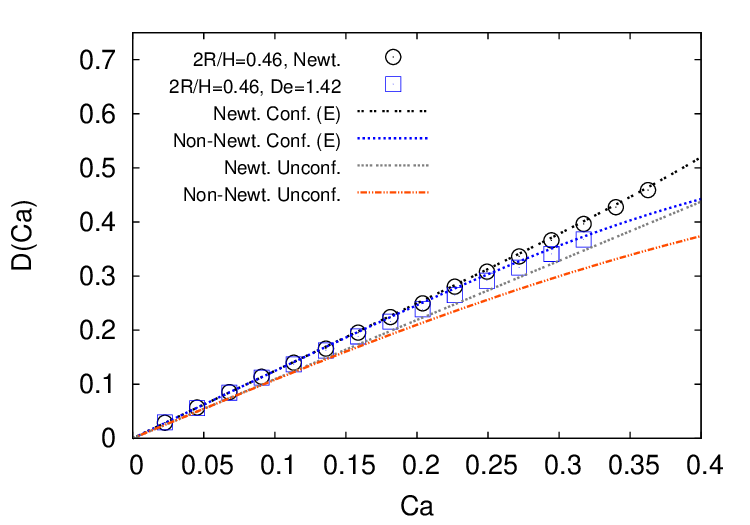}
\includegraphics[scale=0.7]{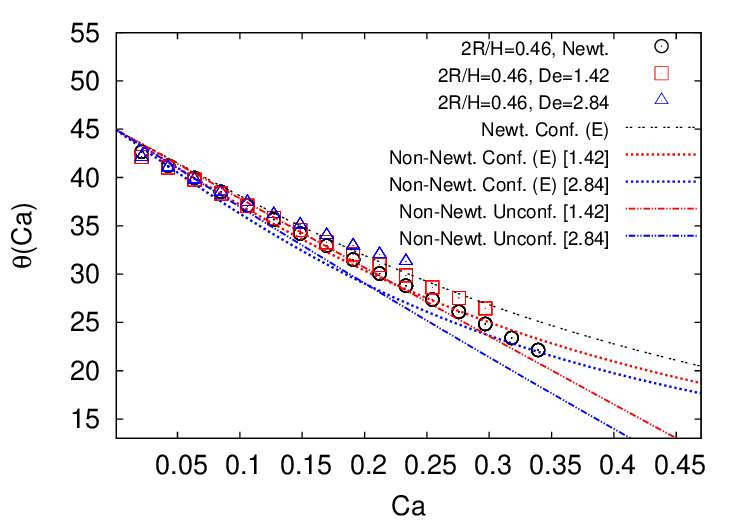}
\caption{We report the steady state deformation parameter $D$ (left panel, see text for details) and the orientation angle (right panel) for a viscoelastic droplet in a  Newtonian matrix under steady shear as a function of the associated Capillary number $Ca$. The viscosity ratio between the droplet phase and the matrix phase is kept fixed to $\lambda=\eta_D/\eta_M=1$, the confinement ratio is $2R/H=0.46$. We consider two relaxation times in the polymer equation (\ref{FENE}), $\tau_P=2000$ lbu and $\tau_P=4000$ lbu, corresponding to  Deborah numbers (based on equation (\ref{DeborahOLDROYD})) $De=1.42$ and $De=2.84$ respectively.  The polymer viscosity is kept fixed to $\eta_P=0.6933$ lbu, corresponding to a polymer concentration of $\eta_P/(\eta_P+\eta_B)=0.4$. With respect to the Newtonian case, deformation is inhibited by viscoelasticity and the numerical results overestimate Greco's prediction for an unconfined non-Newtonian droplet \cite{Greco02} (referred to as ``non-Newtonian unconfined''). As for the orientation, we hardly see any effect. We also report the theoretical predictions of the ``ellipsoidal'' models \cite{Minale08,Minale10} for both Newtonian \cite{Minale08} and non-Newtonian \cite{Minale10} cases (referred to as ``Newtonian confined (E)'' and  ``non-Newtonian confined (E)''). For the non-Newtonian theoretical prediction, smaller angles are related to larger Deborah number. }
\label{fig:deformationNONNEWT1}
\end{center}
\end{figure}

In this section we look at the effects of viscoelasticity in droplet deformation and orientation. We will separately address the importance of matrix viscoelasticity and droplet viscoelasticity, using the proposed methodology described in section \ref{sec:model}, and compare with some of the theoretical predictions available in the literature \cite{Greco02,Minale04,Minale10}. Again, we work with unitary viscosity ratio, defined in terms of the total (fluid+polymer) shear viscosity: $\lambda=(\eta_A+\eta_P)/\eta_B=1$, in case of droplet viscoelasticity; $\lambda=\eta_A/(\eta_B+\eta_P)=1$, in case of matrix viscoelasticity. Viscoelastic effects show up in the droplet deformation and orientation in terms of two dimensionless parameters: the Deborah number, 
\be\label{Deborah}
De=\frac{N_1 R}{2 \sigma_{AB}}\frac{1}{Ca^2}
\ee
where $N_1$ is the first normal stress difference generated in simple shear flow \cite{bird87}, and the ratio $N_2/N_1$ between the second and first normal stress difference \cite{Greco02}. Solving the constitutive equation for steady shear (see section (\ref{simpleshear})), the first normal stress difference for the FENE-P model \cite{bird87,Lindner03} can be computed (see subsection (\ref{simpleshear}) and equation (\ref{lindner2})), while $N_2/N_1=0$. In the Oldroyd-B limit ($L^2 \gg 1$) we can use the asymptotic expansion of the hyperbolic functions and we get  $N_1=2 \eta_P \dot{\gamma}^2 \tau_P$ so that 
\be\label{DeborahOLDROYD}
De=\frac{\tau_P}{\tau_{\mbox{\tiny{em}}}} \frac{\eta_P}{\eta_{M}}
\ee 
showing that $De$ is clearly dependent on the ratio between the polymer relaxation time $\tau_P$ and the emulsion time $\tau_{\mbox{\tiny{em}}}=\frac{R \eta_{M}}{\sigma_{AB}}$, the latter depending on the interface properties (i.e. surface tension). For finite $L^2$, however, we need to use the definition of $De$ based on the first normal stress difference (see section (\ref{simpleshear})). Benchmark tests for the viscoelastic effects will be proposed for both shear-induced droplet deformation and orientation at small $Ca$, although the effects on droplet orientation (especially in a case with matrix viscoelasticity) will be more pronounced. This is because non-Newtonian effects on the drop steady state deformation show up at the second order in $Ca$, while the orientation angle has a correction at first order in $Ca$  \cite{Greco02,Guido11}. In particular, to test both confinement and viscoelastic effects, we will also refer to the model proposed by Minale, Caserta \& Guido \cite{Minale10} for ellipsoidal droplets. Indeed, the aforementioned ellipsoidal models for Newtonian fluids have been recently proposed also for non-Newtonian fluids. In particular, Minale \cite{Minale04} proposed an ellipsoidal model which recovers, in the small $Ca$-limit, the steady state theory developed by Greco \cite{Greco02}. Minale, Caserta \& Guido \cite{Minale10} generalized the work by Minale \cite{Minale04,Minale08} to study the effects of confinement in non-Newtonian systems.\\
We start with the effect of droplet viscoelasticity. For a given confinement ratio, $2R/H=0.46$, in figure \ref{fig:deformationNONNEWT1} we report the steady state droplet deformation and orientation angle. We use the Oldroyd-B model, by choosing a large value of $L^2=10^4$, and  consider two relaxation times in the polymer equation (\ref{FENE}), $\tau_P=2000$ lbu and $\tau_P=4000$ lbu, corresponding to Deborah numbers (based on equation (\ref{DeborahOLDROYD})) $De=1.42$ and $De=2.84$, respectively. The polymer viscosity is kept fixed to $\eta_P=0.6933$ lbu, corresponding to a polymer concentration of $\eta_P/(\eta_A+\eta_P)=0.4$. The deformation computed from the numerical simulations reveals a small effect of viscoelasticity, which is consistent with the theoretical prediction of the model by Minale, Caserta \& Guido \cite{Minale10} (referred to as "non-Newtonian confined (E)''). In particular, with respect to the Newtonian case, deformation is slightly inhibited by viscoelasticity and overestimates Greco's prediction for an unconfined non-Newtonian droplet \cite{Greco02} (referred to as ``non-Newtonian unconfined''). As for the orientation, we again see a small effect. These observations echo other experimental and numerical results present in the literature on the effect of droplet viscoelasticity on deformation and orientation \cite{verhulst09a,verhulst09b,AggarwalSarkar07,AggarwalSarkar08}.

\begin{figure}[t!]
\begin{center}
\includegraphics[scale=0.7]{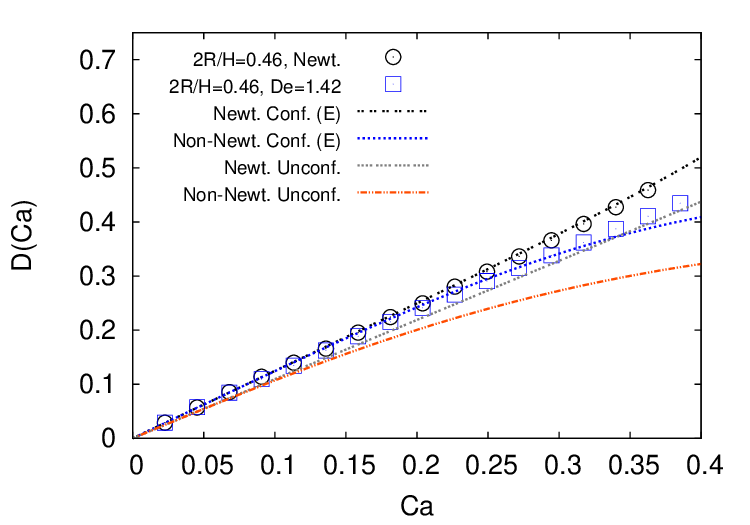}
\includegraphics[scale=0.7]{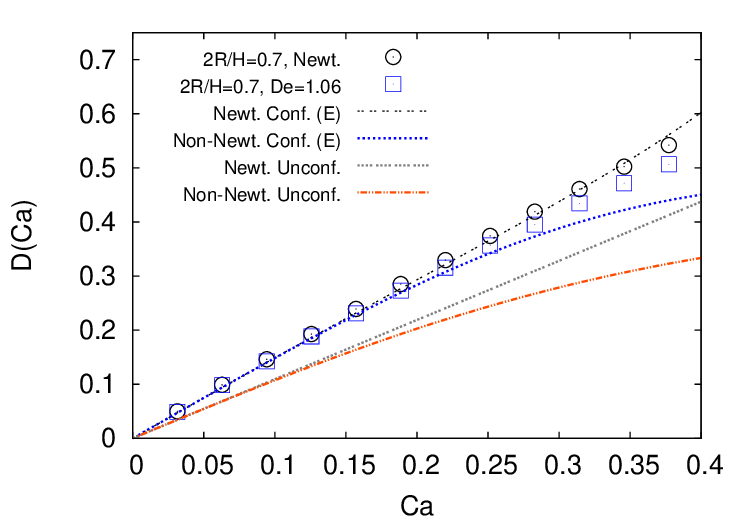}
\caption{We report the steady state deformation parameter $D$ (see text for details) for a Newtonian droplet in a viscoelastic matrix under steady shear as a function of the Capillary number $Ca$. The viscosity ratio between the droplet phase and the matrix phase is kept fixed to $\lambda=\eta_D/\eta_M=1$. Two different confinement ratios are considered: $2R/H=0.46$ (left panel) and $2R/H=0.7$ (right panel). Again, as already done for the data of figure \ref{fig:deformationNONNEWT1}, we choose a large value of the finite extensibility parameter $L^2=10^4$, and  consider a relaxation time in the polymer equation (\ref{FENE}) $\tau_P=2000$ lbu. The corresponding Deborah numbers depend on the droplet radius, based on equation (\ref{DeborahOLDROYD}): $De=1.42$ for $2R/H=0.46$ and $De=1.06$ for $2R/H=0.7$. The polymer viscosity is kept fixed to $\eta_P=0.6933$ lbu, corresponding to a polymer concentration of $\eta_P/(\eta_P+\eta_B)=0.4$.  The numerical results overestimate Greco's prediction for an unconfined non-Newtonian droplet \cite{Greco02} (referred to as ``non-Newtonian unconfined''). We also report the prediction of ``ellipsoidal'' models \cite{Minale08,Minale10} for both Newtonian \cite{Minale08} and non-Newtonian \cite{Minale10} cases (referred to as ``Newtonian confined (E)'' and ``non-Newtonian confined (E)'').}
\label{fig:deformationNONNEWT2}
\end{center}
\end{figure}

\begin{figure}[tbp]
\begin{center}
\includegraphics[scale=0.7]{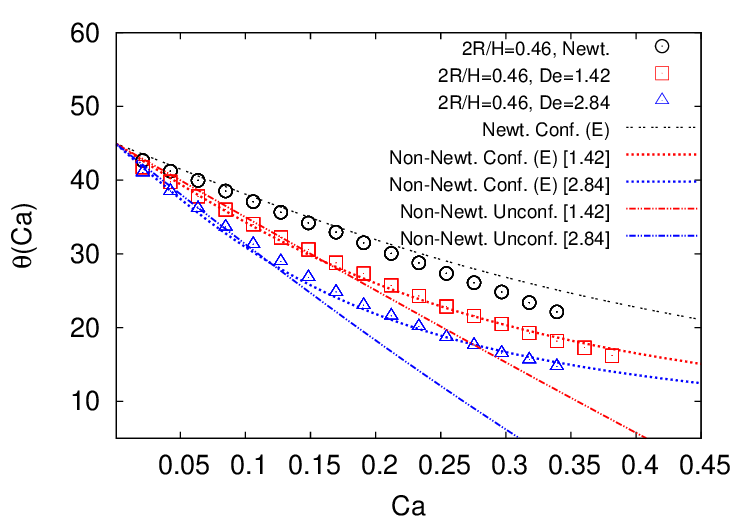}
\includegraphics[scale=0.7]{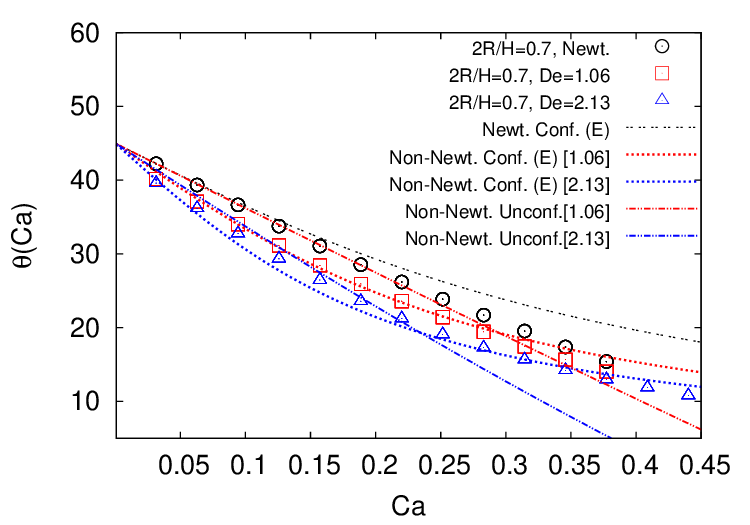}
\caption{We report the steady state orientation angle for a Newtonian droplet in a viscoelastic matrix under steady shear as a function of the Capillary number $Ca$. The viscosity ratio between the droplet phase and the matrix phase is kept fixed to $\lambda=\eta_D/\eta_M=1$. Two different confinement ratios are considered: $2R/H=0.46$ (left panel) and $2R/H=0.7$ (right panel). Data are the same as those of figure \ref{fig:deformationNONNEWT2}, plus some other data obtained by increasing the relaxation time $\tau_P$ in equation (\ref{FENE}). For a given $Ca$, the numerical results overestimate Greco's prediction for an unconfined non-Newtonian droplets \cite{Greco02} (referred to as ``non-Newtonian unconfined''). We also report the theoretical predictions of the ``ellipsoidal'' models \cite{Minale08,Minale10} for both Newtonian and non-Newtonian cases (referred to as ``Newtonian confined (E)'' and ``non-Newtonian confined (E)''). For the non-Newtonian theoretical prediction, smaller angles are related to larger Deborah number. \label{fig:orientationNONNEWT}}
\end{center}
\end{figure}

\begin{figure}[hp!]
{
\includegraphics[scale=0.15]{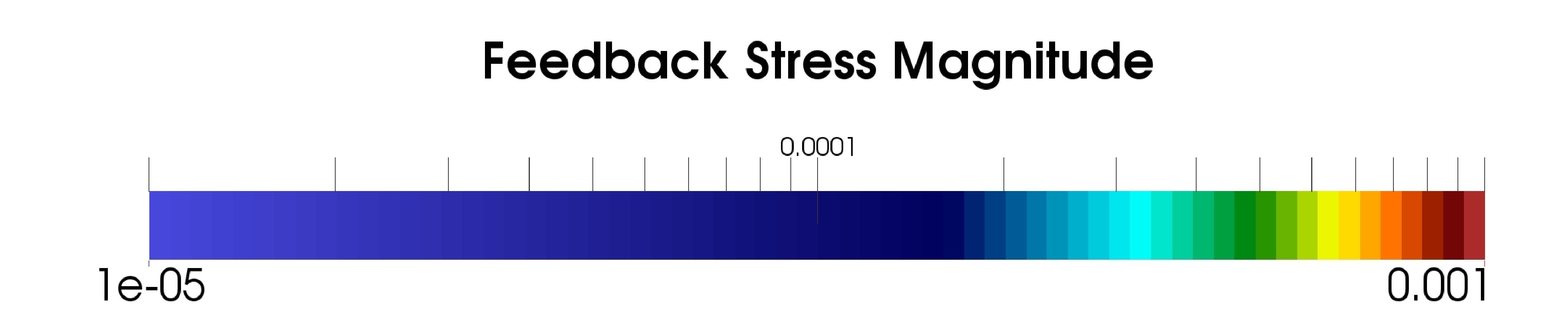}
}
\\
\subfigure[\,\, 2R/H=0.46, Ca= 0.17 , De=0.71]
{
\includegraphics[scale=0.1]{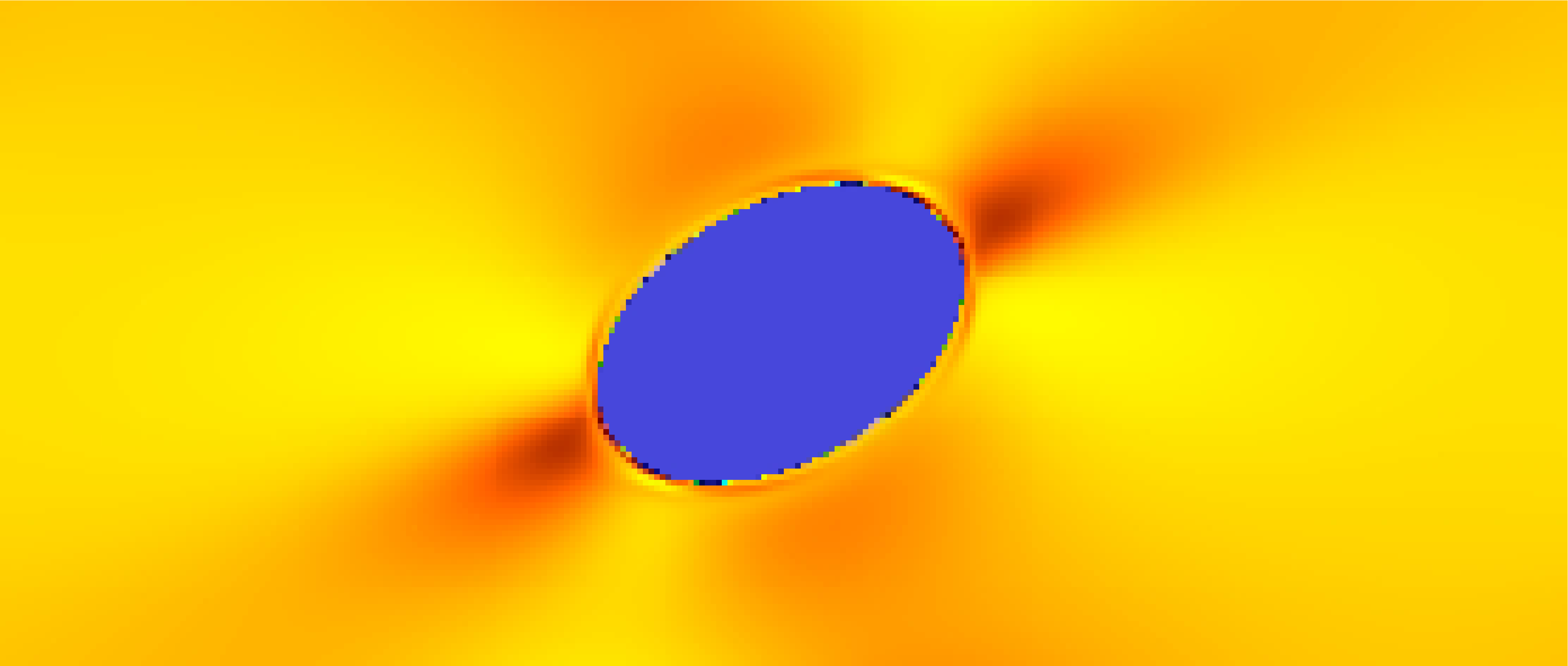}
\label{fig:8d}
}    
\subfigure[\,\, 2R/H=0.7, Ca= 0.23 , De=0.53]
{
\includegraphics[scale=0.1]{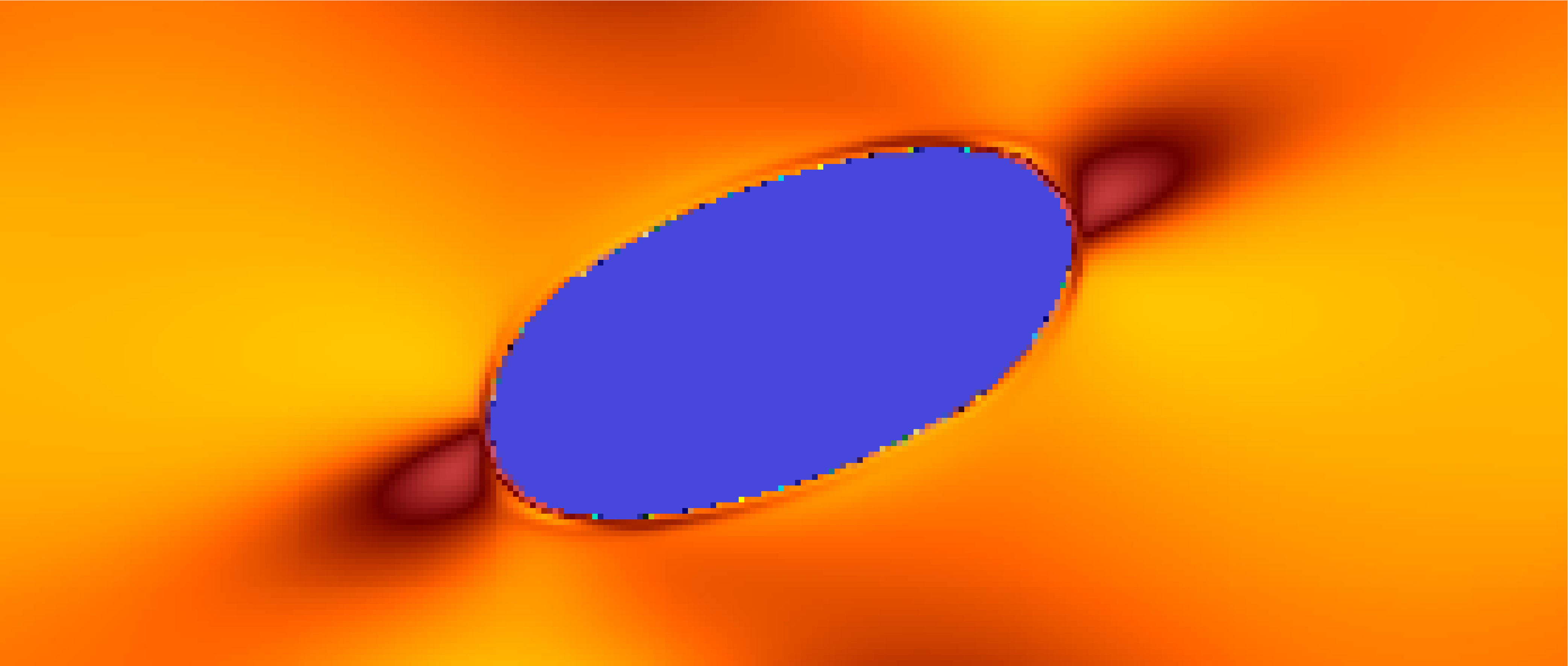}
\label{fig:8e}
}    
\\
\subfigure[\,\, 2R/H=0.46, Ca= 0.17 , De=1.42]
{
\includegraphics[scale=0.1]{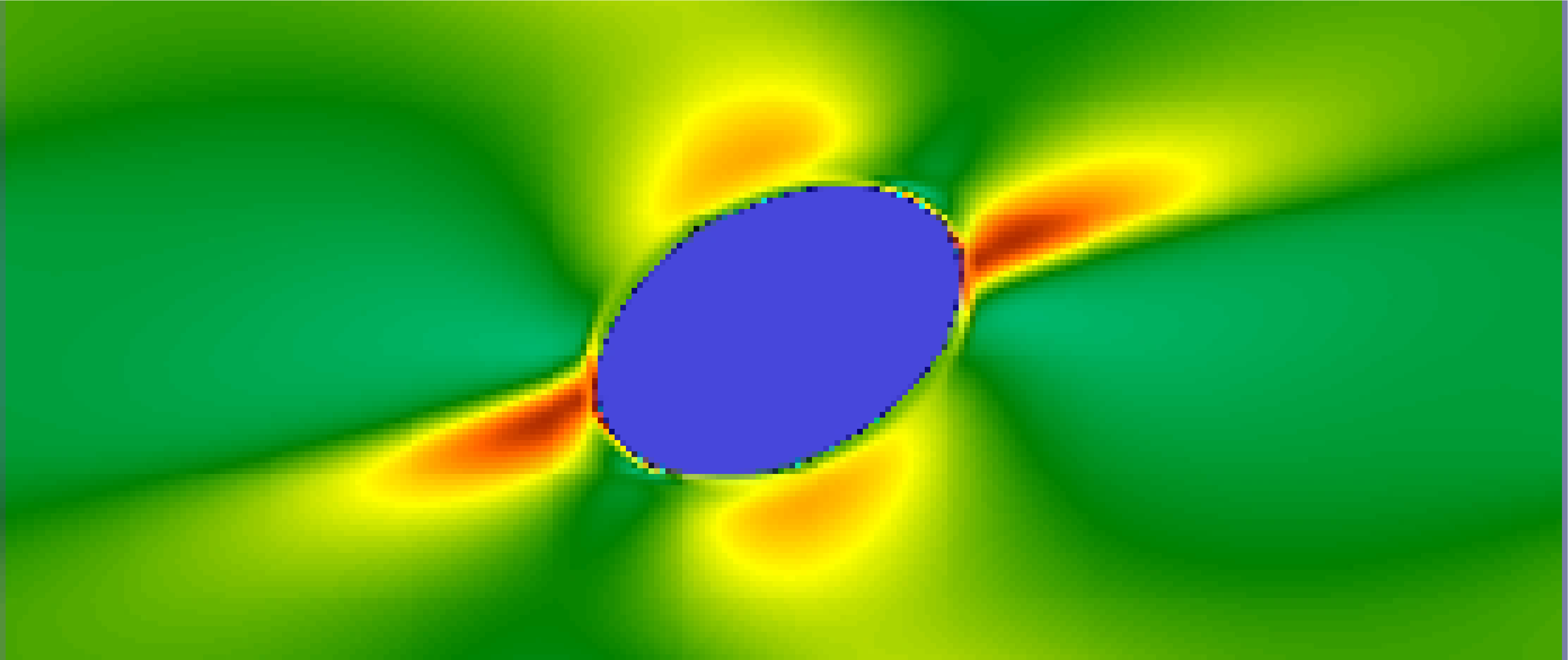}
\label{fig:8d}
}    
\subfigure[\,\, 2R/H=0.7, Ca= 0.23 , De=1.06]
{
\includegraphics[scale=0.1]{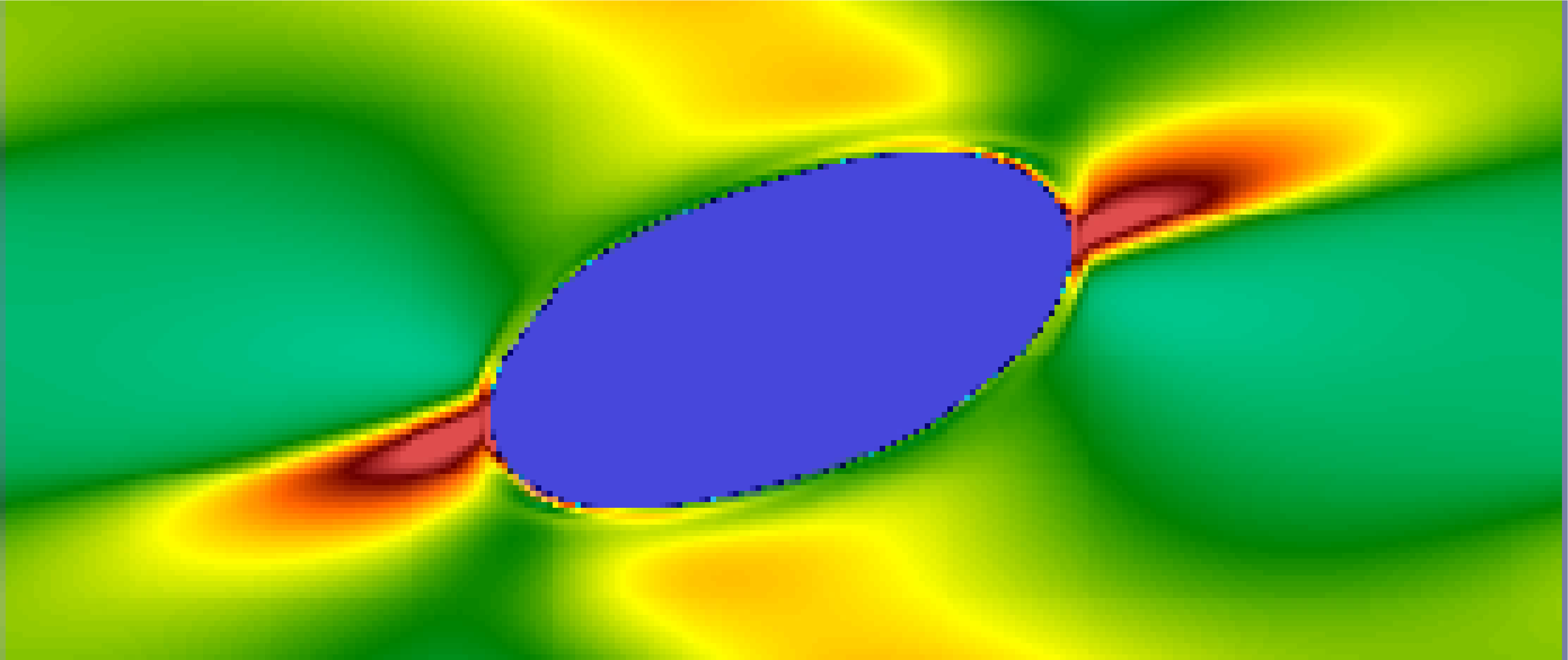}
\label{fig:8e}
}    
\\
\subfigure[\,\, 2R/H=0.46, Ca= 0.17 , De=2.84]
{
\includegraphics[scale=0.1]{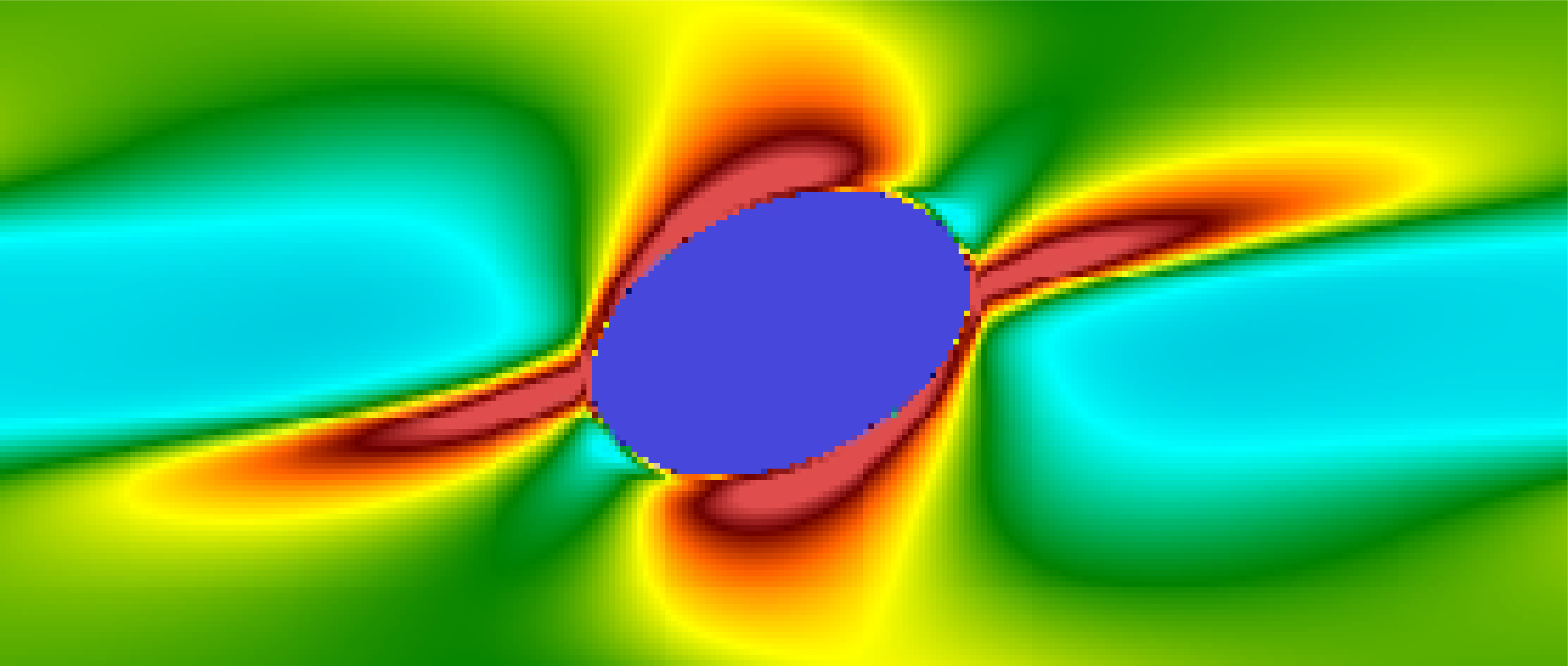}
\label{fig:8d}
}    
\subfigure[\,\, 2R/H=0.7, Ca= 0.23 , De=2.13]
{
\includegraphics[scale=0.1]{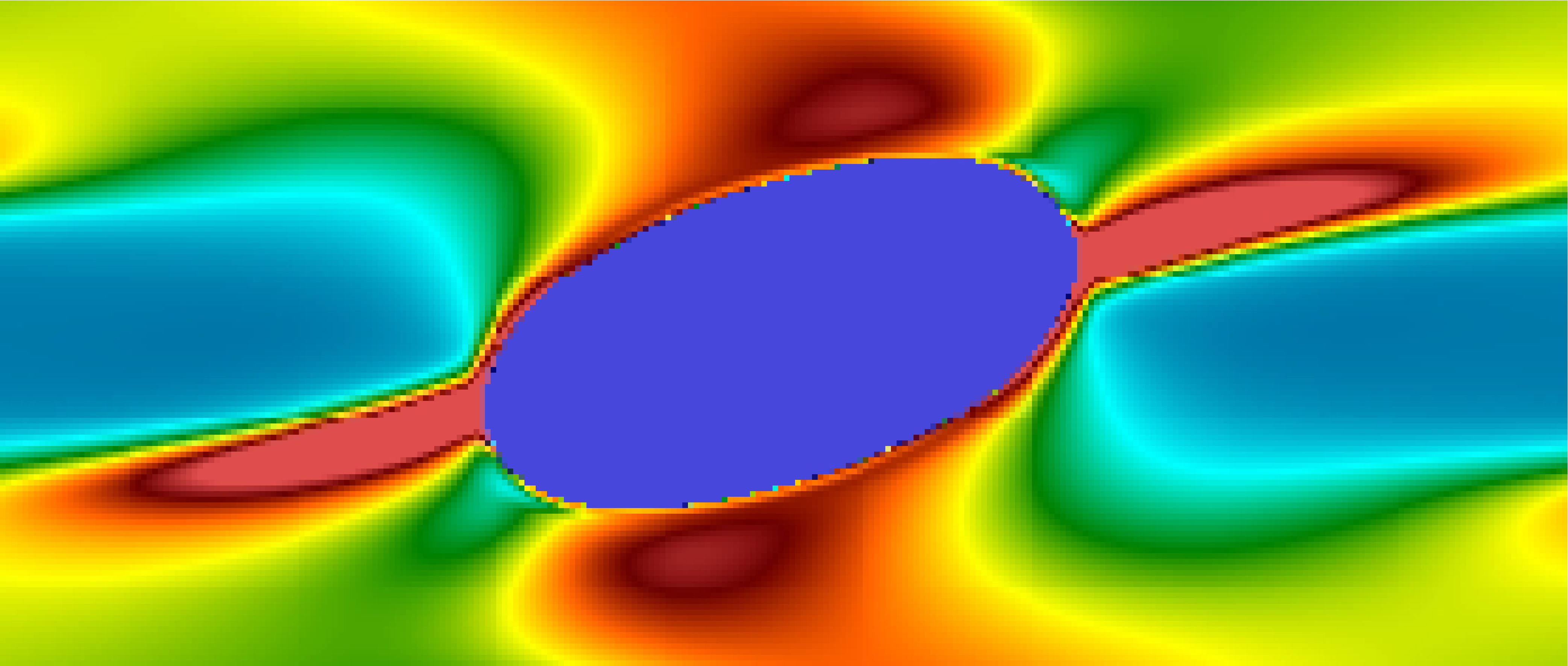}
\label{fig:8e}
}    
\caption{We report the steady state snapshots of the polymer feedback stress in equation (\ref{NS}) for the cases studied in figures \ref{fig:deformationNONNEWT2} and \ref{fig:orientationNONNEWT} in the plane $z=H/2$. Results are obtained for the same wall velocity, $U_w=\pm 0.02$ lbu, the same finite extensibility parameter $L^2=10^4$, and considering three different relaxation times in the polymer equation (\ref{FENE}), $\tau_P=1000, 2000, 4000$ lbu. The corresponding Deborah numbers depend on the droplet radius, based on equation (\ref{DeborahOLDROYD}): $De=0.71, 1.42, 2.84$ for $2R/H=0.46$ and $De=0.53, 1.06,2.13$ for $2R/H=0.7$.  \label{fig:STRESSCOREO}}.
\end{figure}

We next look at the effect of matrix viscoelasticity, figures \ref{fig:deformationNONNEWT2} and \ref{fig:orientationNONNEWT}. In figure \ref{fig:deformationNONNEWT2} we report the steady state droplet deformation for two different confinement ratios: $2R/H=0.46$ (left panel) and $2R/H=0.7$ (right panel). Again, we choose a large value of $L^2=10^4$, and  consider a relaxation time $\tau_P=2000$ lbu in the polymer equation (\ref{FENE}),  corresponding to different Deborah numbers, depending on the droplet radius (see equation (\ref{DeborahOLDROYD})): $De=1.42$ for $2R/H=0.46$ and $De=1.06$ for $2R/H=0.7$. The polymer viscosity is kept fixed to $\eta_P=0.6933$ lbu, corresponding to a polymer concentration of $\eta_P/(\eta_P+\eta_B)=0.4$. In both cases, matrix viscoelasticity inhibits droplet deformation with respect to the corresponding Newtonian cases. Also, the unconfined theory by Greco \cite{Greco02} underestimates the deformation, and the mismatch is larger with the larger confinement ratio, as one would have expected since the theory of Greco does not take into account confinement. The model by Minale, Caserta \& Guido \cite{Minale10} follows the numerical data with a mismatch emerging at large $Ca$ for the larger confinement ratio: most probably this is due to the fact that confinement starts to act in promoting deformation with shapes departing from an ellipsoid \cite{Sibillo06}. A non trivial interplay between confinement and viscoelasticity is also visible from figure \ref{fig:STRESSCOREO}, where we report the steady state snapshots for the polymer feedback stress of equation (\ref{NS}) for the cases studied in figures \ref{fig:deformationNONNEWT2} and \ref{fig:orientationNONNEWT}. In figure \ref{fig:orientationNONNEWT} we report the orientation angle for the same cases studied in figure \ref{fig:deformationNONNEWT2}. The effect of viscoelasticity is now much more visible, if compared with the case of droplet viscoelasticity reported in figure \ref{fig:deformationNONNEWT1}. We also analyze the effect of an increase of the relaxation time $\tau_P$ in equation (\ref{FENE}) for both the confinement ratios studied, which translates in a larger Deborah number. The change in the orientation angle for the Newtonian cases is linear in $Ca$ up to the largest $Ca$ considered, which is consistent with the linearity of the deformation discussed in figure \ref{fig:deformationNewtonian}. This generates a mismatch with the corresponding Ellipsoidal model predictions \cite{Minale08}: just to give some quantitative numbers, for a Capillary number $Ca=0.35$, there is a mismatch of $2-3^{\circ}$  in the smaller confinement ratio, which becomes roughly doubled (i.e. $5-6^{\circ}$) for the larger confinement ratio.  The orientation angle in the non-Newtonian cases, instead, is better captured by the ellipsoidal model by Minale, Caserta \& Guido \cite{Minale10}.  Overall, in both the Newtonian and non-Newtonian cases, the mismatch between the numerical results and the prediction of the ellipsoidal models is more pronounced at large confinement ratios (right panel of figure \ref{fig:orientationNONNEWT}), an observation that echoes the discussion done for the data of figure \ref{fig:deformationNONNEWT2}.

\begin{figure}[t!]
\begin{center}
\includegraphics[scale=0.7]{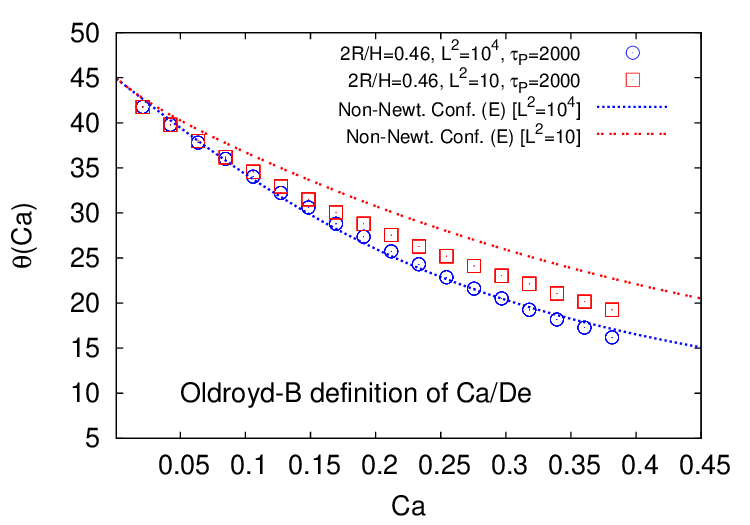}
\includegraphics[scale=0.7]{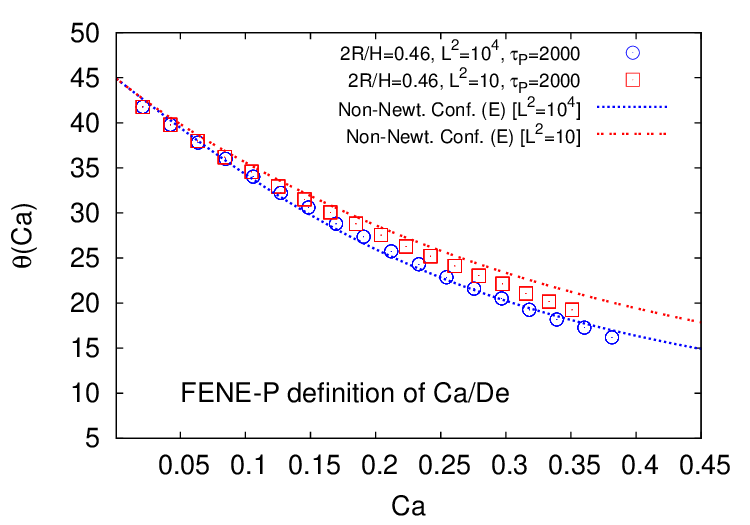}
\caption{Left Panel: we report the steady state orientation angle for a Newtonian droplet in a viscoelastic matrix under steady shear as a function of the Capillary number $Ca$. For a given confinement ratio $2R/H=0.46$ and $\tau_P=2000$ lbu in equation (\ref{FENE}), we have repeated the numerical simulations described in the left panel of figure \ref{fig:deformationNONNEWT2} for a  finite extensibility parameter $L^2=10$. We have used the definition of Deborah number based on equation (\ref{DeborahOLDROYD}) and a shear independent matrix viscosity in equation (\ref{capillary}) to compute $Ca$. These choices are appropriate only in the Oldroyd-B limit ($L^2 \gg 1$), hence referred to as ``Oldroyd-B definition''. Right Panel: we report the same data of the left panel by changing the definition of Capillary number in equation (\ref{capillary}), based on the thinning effects analyzed in section (\ref{sec:dilutesuspensions}), and changing the definition of the Deborah number which is now computed according to equation (\ref{Deborah}). This is referred to as ``FENE-P definition''. For the non-Newtonian theoretical prediction, smaller angles are related to larger $L^2$. Steady state snapshots of the polymer feedback stress in equation (\ref{NS}) for some of these cases are reported in figure \ref{fig:L2COREO}. \label{fig:ORIENTATIONL2}}
\end{center}
\end{figure}

\begin{figure}[h!]
{
\includegraphics[scale=0.15]{BARRA.eps}
}
\\
\subfigure[\,\, 2R/H=0.46, $L^2=10^4$, $\tau_P=2000$]
{
\includegraphics[scale=0.08]{RW_0.46_taup_2000_matrix.eps}
\label{fig:8d}
}    
\subfigure[\,\, 2R/H=0.46, $L^2=10$, $\tau_P=2000$]
{
\includegraphics[scale=0.08]{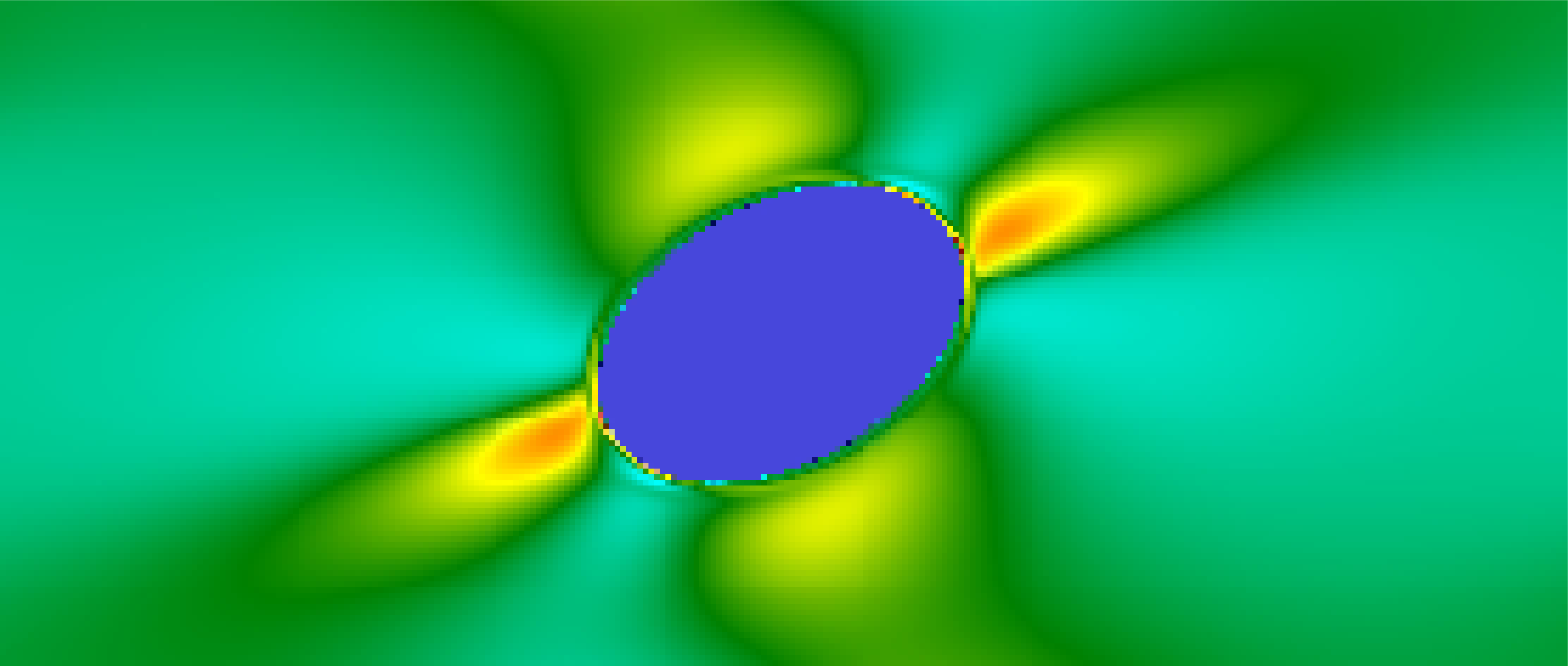}
\label{fig:8e}
}    
\caption{We report the steady state snapshots of the polymer feedback stress in equation (\ref{NS}) for the cases studied in figure \ref{fig:ORIENTATIONL2} in the plane $z=H/2$. Results are obtained for the same wall velocity, $U_w=\pm 0.02$ lbu, the same relaxation time $\tau_P=2000$ lbu in the polymer equation (\ref{FENE}), and different finite extensibility parameters $L^2=10$ and $L^2=10^4$. The corresponding Deborah numbers depend on the droplet radius, based on equation (\ref{Deborah}). In both cases, the polymer viscosity is kept fixed to $\eta_P=0.6933$ lbu, corresponding to a polymer concentration of $\eta_P/(\eta_P+\eta_B)=0.4$, but the case with $L^2=10$ has thinning effects in regions with large shears (see also section (\ref{sec:dilutesuspensions})). \label{fig:L2COREO}}.
\end{figure}

Finally, we want to address and test the importance of the finite extensibility parameter in the polymer equation (\ref{FENE}). For a given confinement ratio $2R/H=0.46$ and $\tau_P=2000$ lbu in equation (\ref{FENE}), we have repeated the numerical simulations described in the left panel of figure \ref{fig:deformationNONNEWT2} for a finite extensibility parameter $L^2=10$. As $L$ decreases, the polymer dumbell becomes less extensible and the maximum level of stress attainable is reduced. There are some consequences. First, we cannot rely on equation (\ref{DeborahOLDROYD}) to define the Deborah number, which strictly holds only in the large-$L^2$ limit. Second, at large shears, the model exhibits thinning effects, as predicted and verified in subsection (\ref{simpleshear}), and the definition of the Capillary number (\ref{capillary}) given in terms of the matrix viscosity has to be changed to include such effects.  Indeed, by using the definition of the Deborah number given in equation (\ref{DeborahOLDROYD}) and a shear independent matrix viscosity in equation (\ref{capillary}) in the theoretical models, the agreement between the numerical results and the theory deteriorates (see left panel of figure \ref{fig:ORIENTATIONL2}), whereas the large-$L^2$ case was well in agreement. In the right panel of figure \ref{fig:ORIENTATIONL2} we report the same data, by changing: (i) the definition of Capillary in equation (\ref{capillary}), based on the thinning effects analyzed in subsection (\ref{simpleshear}); (ii) the definition of the Deborah number, which is now computed according to equation (\ref{Deborah}), with the first normal stress difference given in (\ref{lindner2}). As one can see the agreement gets better, especially at small $Ca$. For completeness, in figure \ref{fig:L2COREO}, we report the steady state snapshots for the polymer feedback stress of equation (\ref{NS}) for the cases studied in figure \ref{fig:ORIENTATIONL2}. Results reported in figure \ref{fig:ORIENTATIONL2} are surely motivating for future theoretical studies. Indeed, it is by no means proved that the theoretical models used \cite{Minale10} can work for a shear-dependent viscosity (which holds for the FENE-P).  Figure \ref{fig:ORIENTATIONL2} is giving (numerical) evidence that the viscoelastic effects of the FENE-P model can also be embedded in such theoretical models; however, before proceeding with further comparisons, we feel that a proper theoretical background needs to be developed first.

\section{Conclusions}
We have proposed numerical simulations of viscoelastic fluids based on a hybrid algorithm combining lattice-Boltzmann models (LBM) and Finite Differences (FD) schemes, the former used to model the macroscopic hydrodynamic equations, and the latter used to model the kinetics of polymers using the constitutive equations for finitely extensible non-linear elastic dumbells with Peterlin's closure (FENE-P). We have first benchmarked the numerical scheme with the characterization of the rheological properties of a dilute homogeneous solution under steady shear, steady elongational flows, oscillatory flows and transient shear. We then continued to study the model in presence of non-ideal multicomponent interfaces, where immiscibility is introduced in the LBM description using the ``Shan-Chen'' interaction model \cite{SC93,SC94,CHEM09}. We have characterized the effect of viscoelasticity in droplet deformation under steady shear, by comparing the results of numerical simulations with available theoretical models in the literature \cite{Greco02,Minale04,ShapiraHaber90,Minale08,Minale10,Minale10review,Taylor34}. Overall, the numerical simulations well capture both the effects of confinement and viscoelasticity, thus exploring problems where the capabilities of LBM were never quantified before. Even if we focused on a unitary total (Newtonian fluid+polymer) viscosity ratio, the numerical algorithm can simulate viscosity ratios different from $1$ as well, although we think that the latter cannot easily be pushed much below $0.1$ and much above $10$. Based on the total shear viscosity, we actually show in this paper that there is a good matching between the analytical solutions and the numerical results for those cases where the viscosity ratio between the two Newtonian phases is between $0.66$ and $1.6$, while keeping the total (Newtonian fluid+Polymer) viscosity ratio equal to $1$.  We think the good matching is possible only because the ``bare'' Newtonian solution is recovered, therefore lending support to the validity of the algorithm in simulating viscosity ratios different from $1$.
As an upgrade of complexity, it would be extremely interesting to study time-dependent situations \cite{verhulst09b,Cardinaels10}, other flows in confined geometries \cite{Arratia,Garstecky} and problems where droplet break-up is involved \cite{Cardinaelsetal11b,GuptaSbragaglia}. Complementing these kind of experimental results with the help of numerical simulations would be of extreme interest. Simulations provide easy access to quantities such as drop deformation and orientation as well as the velocity flow field, pressure field, and polymers feedback stresses, inside and outside the droplet. They can be therefore useful to perform in-silico  comparative studies, at changing the model parameters, to shed lights on the complex properties of viscoelastic flows in confined geometries. 

\section{Acknowledgements}

The authors kindly acknowledge funding from the European Research Council under the European Community's Seventh Framework Programme (FP7/2007-2013)/ERC Grant Agreement no[279004]. We also acknowledge L. Biferale and R. Benzi for useful discussions. A. Gupta acknowledges R. Pandit and S. S. Ray for useful discussions during his visits to IISc Bangalore (February 2014) and ICTS-TIFR Bangalore (October 2013).

\appendix

\section{Sensitivity with respect to a change in the resolution and model parameters used}\label{AppendixA}

The convergence towards the sharp-interface limit of hydrodynamics is one of the crucial issues in diffuse interface models \cite{VanDerSman08,Komrakovaa13,casciola13,Anderson,Lowengrub}. In the present work, the resolution used, the interface thickness, the mobility were all empirically tuned in order to match the analytical predictions of sharp-interface (Newtonian) hydrodynamics. In this Appendix we provide evidence that the chosen parameters lie in a range of values where the hydrodynamic solution is indeed well recovered. The reference numerical data are those analyzed in the right panel of figure \ref{fig:deformationNewtonian} for $2R/H=0.46$. The sensitivity with respect to a change in the resolution used is tested by keeping all the parameters fixed, except the wall-to-wall gap $H$ and the radius $R$, which are changed in the ranges $64 \le H \le 176$ lattice cells and $12 \le R \le 40$ lattice cells, in such a way to keep fixed the confinement ratio to $2R/H=0.46$. All the numerical simulations performed match very well with the theoretical prediction of Shapira \& Haber for a confined Newtonian droplet \cite{ShapiraHaber90} (referred to as ``Newtonian confined''). Even the simulations with the smallest radius analyzed ($R=12$ lattice cells) are well in agreement with the theoretical predictions, a fact that is also acknowledged in other publications using the ``Shan-Chen'' interaction model \cite{Onishi2}.

\begin{figure}[tbp]
\begin{center}
\includegraphics[scale=0.7]{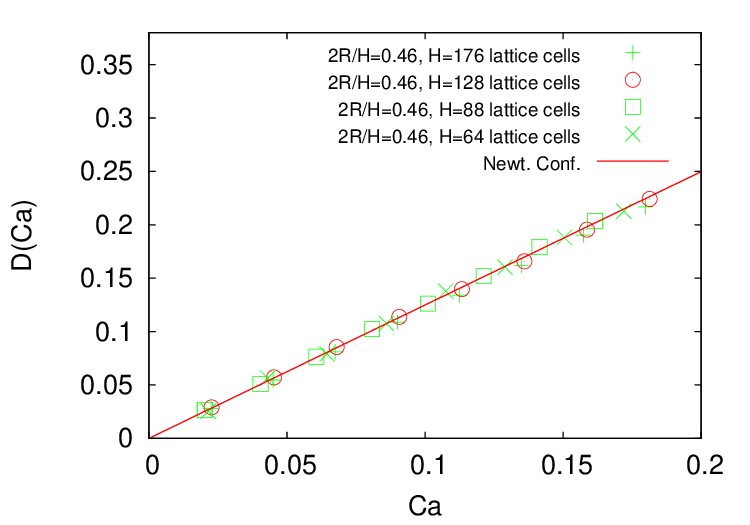}
\caption{We report the steady state deformation parameter $D$ for a Newtonian droplet in a Newtonian matrix under steady shear as a function of the associated Capillary number $Ca$. We start from the data reported the right panel of figure \ref{fig:deformationNewtonian} with $2R/H=0.46$. We vary the wall-to-wall gap $H$, by keeping the confinement ratio fixed to $2R/H=0.46$. All the other parameters are kept fixed. The theoretical prediction of Shapira \& Haber for a confined Newtonian droplet \cite{ShapiraHaber90} (referred to as ``Newtonian confined'') is also reported.}\label{fig:resolution}
\end{center}
\end{figure}

\begin{figure}[tbp]
\includegraphics[scale=0.7]{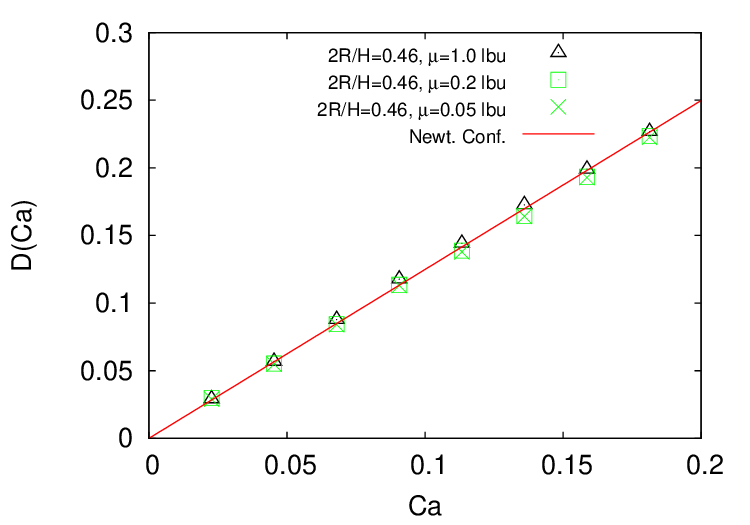}
\includegraphics[scale=0.7]{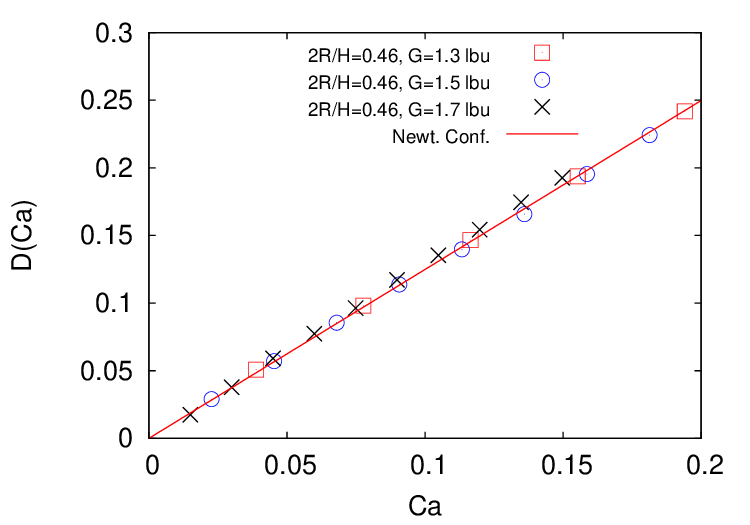}
\caption{We report the steady state deformation parameter $D$ for a Newtonian droplet in a Newtonian matrix under steady shear as a function of the associated Capillary number $Ca$. We start from the data reported the right panel of figure \ref{fig:deformationNewtonian} with $2R/H=0.46$. In a series of numerical simulations, we change the mobility $\mu$ in \eqref{DIFFUSIONCURRENT}-\eqref{TRANSPORTCOEFF} by keeping all the other parameters fixed (left panel). In another set of simulations, we change the interaction parameter ${\cal G}$ in \eqref{eq:SCforce}, thus obtaining various situations with different interface widths (right panel, see also text for details). The theoretical prediction of Shapira \& Haber for a confined Newtonian droplet \cite{ShapiraHaber90} (referred to as ``Newtonian confined'') is also reported.}
\label{fig:sensitivity}
\end{figure}

We next continue by performing numerical simulations to test the sensitivity with respect to a change in the mobility $\mu$ (see \eqref{DIFFUSIONCURRENT}-\eqref{TRANSPORTCOEFF}) and in the interface width. In both cases, again, the reference numerical data are those analyzed in the right panel of figure \ref{fig:deformationNewtonian} for $2R/H=0.46$, corresponding to a mobility $\mu=0.5$ lbu and interface width approximately equal to $5$ lattice cells. In a series of numerical simulations, we change the mobility in the range $0.05 \le \mu \le 1.0$ lbu, by keeping all the other parameters unchanged. Results are reported in the left panel of figure \ref{fig:sensitivity}, showing no remarkable sensitivity, at least as far as the deformation parameter is concerned. In a second set of numerical simulations, we keep the mobility fixed to $\mu=0.5$ lbu and change the interface width: the interaction parameter is changed in the range $1.3 \le {\cal G} \le 1.7$ lbu at fixed total average density, resulting in surface tensions varying in the range $ 0.05 \le \sigma_{AB} \le 0.14$ lbu. The interface widths are changed in a range between $3$ lattice cells and $8$ lattice cells approximately (wider interfaces are obtained with smaller ${\cal G}$). The associated bulk densities change in the range $1.9 \le \rho_A \le 2.15$ lbu and $0.05 \le \rho_B \le 0.2$ lbu in the $A$-rich region. For each value of ${\cal G}$, we define the Capillary number according to \eqref{capillary}, dependently on the value of the surface tension. The results for the deformation parameter $D$ as a function of the Capillary number are reported in the right panel of figure \ref{fig:sensitivity}, confirming that the parameters used in our study lie in a range where the convergence towards the sharp-interface limit of hydrodynamics is well achieved.

\begin{figure}[tbp]
\begin{center}
\includegraphics[scale=0.7]{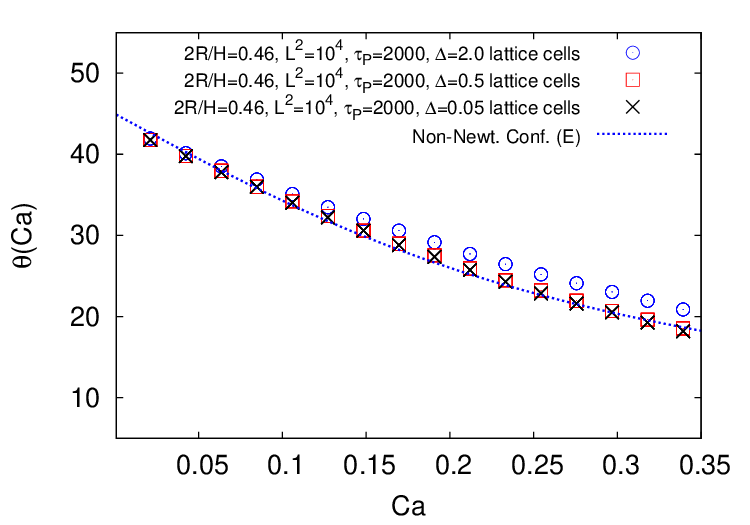}
\caption{Sensitivity of the numerical results with respect to a change in the smoothing parameter $\Delta$ for the viscoelastic properties \eqref{smoothingvisco}. We use the data analyzed in the left panel of figure \ref{fig:orientationNONNEWT} corresponding to the Deborah number $De=1.42$. We report the steady state orientation angle for a Newtonian droplet in a viscoelastic matrix under steady shear as a function of the Capillary number $Ca$, and we change the smoothing parameter $\Delta$ for the viscoelastic properties \eqref{smoothingvisco} in the range $0.01 \le \Delta \le 2$ lattice cells. We also report the theoretical prediction of the ``ellipsoidal'' models \cite{Minale08,Minale10} (referred to as ``non-Newtonian confined (E)'').}
\label{fig:sensitivity_angle}
\end{center}
\end{figure}

We finally address the importance of the smoothing parameter $\Delta$ for the viscoelastic properties \eqref{smoothingvisco}. We choose the data analyzed in the left panel of figure \ref{fig:orientationNONNEWT} for the Deborah number $De=1.42$. The smoothing parameter is changed in the range $0.01 \le \Delta \le 2$ lattice cells and results are reported in figure \ref{fig:sensitivity_angle}. As expected, for values of $\Delta$ below 1 lattice cell, the results are all well in agreement withe the reference theory of sharp-interface hydrodynamics. Deviations start to emerge when the smoothing parameter is above a lattice cell.

\section{Comparison with the Model by Onishi {\it et. al.}}\label{AppendixB}

\begin{figure}[tbp]
\includegraphics[scale=0.7]{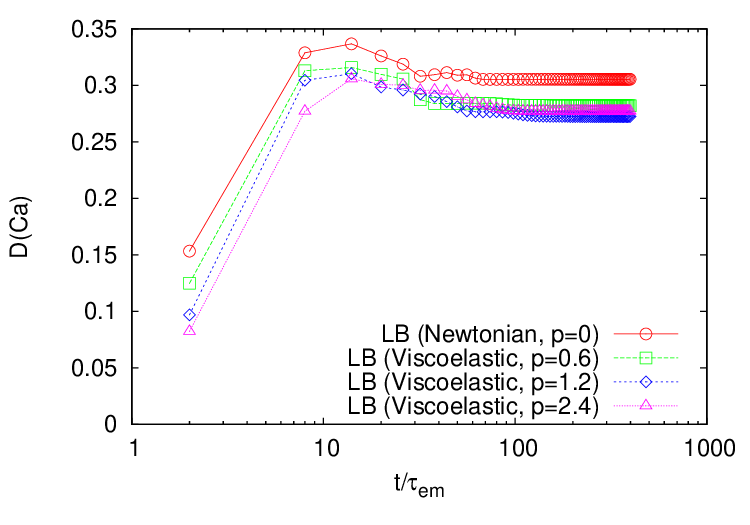}
\includegraphics[scale=0.7]{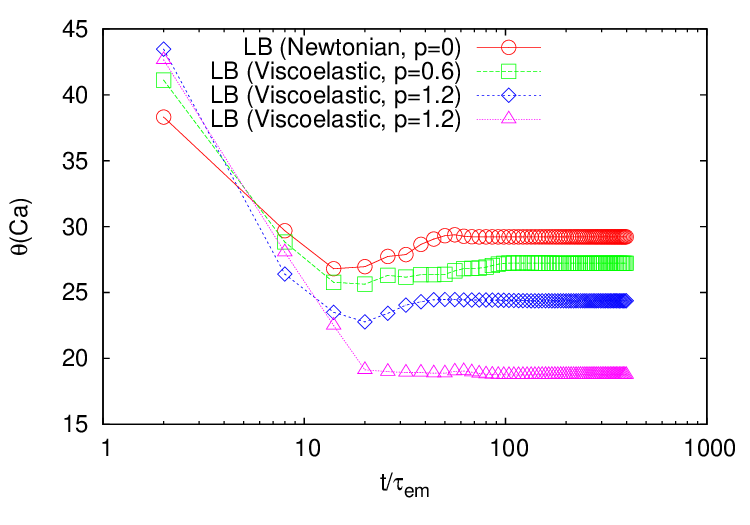}
\caption{Comparisons with the results of Onishi {\it et al.} \cite{Onishi2} for the temporal evolution of Taylor's deformation parameter and orientation angle at fixed Capillary number $Ca=0.26$. Both viscoelastic and Newtonian cases are considered (see text for details). To be noted that the Deborah number $De$ is denoted with $p$ in \cite{Onishi2}: we therefore decided to use $p$ to better (visually) establish a link with the results of \cite{Onishi2}.}
\label{fig:onishi1}
\end{figure}

In this Appendix we compare the results of our model with those of Onishi {\it et al.} \cite{Onishi2}. The two approaches are intrinsically different with regard to the modelling of the polymer dynamics: Onishi {\it et al.} use an approach based on the Fokker-Planck equation (simulated with LBM), whereas we directly model the conformation tensor dynamics (with FD), which comes from a proper closure of the Fokker-Planck equation \cite{bird87,Herrchen97}. Comparing the two theoretical formulations is outside the scope of our paper, and surely addressed in many other dedicated works \cite{bird87,Herrchen97}. The comparison between the two models can be fairly addressed, at least as far as it concerns the solvent part of the model, which is done with LBM in both cases. We do not propose anything new in this direction, since we use the MRT, whose advantages with respect to the single time BGK relaxation approximation (used by Onishi {\it et al.} \cite{Onishi2}) are well known from the literature \cite{Yu,DHumieres02}. These facts said, from the information provided in the paper by Onishi {\it et al.} \cite{Onishi2}, we could run numerical simulations to compare with the results there reported. We perform numerical simulations in three dimensional domains with $L_x \times H \times H = 128 \times 64 \times 64$ lattice cells and droplet radius $R=12$ lattice cells, which is the same resolution used by Onishi {\it et al.} in their paper \cite{Onishi2}. Exactly as in \cite{Onishi2}, we prepare four fluids with viscoelasticity in the matrix phase: the matrix viscosity is kept the same, $\eta_M=\eta_B+\eta_P=2$ lbu, but different viscoelastic properties are considered: $\eta_P/\eta_M=0$ and $De=0.0$ (run M1 in \cite{Onishi2}, the Newtonian case); $\eta_P/\eta_M=0.25$ and $De=0.6$ (run M2 in \cite{Onishi2});  $\beta=\eta_P/\eta_M=0.5$ and $De=1.2$ (run M3 in \cite{Onishi2}); $\beta=\eta_P/\eta_M=0.5$ and $De=2.4$ (run M4 in \cite{Onishi2}). In all cases, the mobility in \eqref{TRANSPORTCOEFF} is set to $\mu=0.5$ lbu. In figure \ref{fig:onishi1}, similarly to figure 3 of Onishi {\it et al.} \cite{Onishi2}, we begin with the presentation of the temporal evolution of Taylor's deformation parameter and orientation angle obtained in runs M1-M4 for a fixed Capillary number $Ca=0.26$. To be noted that the Deborah number $De$ is denoted with $p$ in \cite{Onishi2}: we therefore decided to use $p$ to better (visually) establish a link with the results of \cite{Onishi2}. In agreement with \cite{Onishi2}, there is no remarkable difference in the approach to steady state, though overshoots are a bit more pronounced in our case. Note that we have made time dimensionless with respect to the droplet emulsion time $\tau_{\mbox{\tiny{em}}}=\frac{R \eta_{M}}{\sigma_{AB}}$, whereas it is not clearly stated what is the characteristic time scale used by the authors in \cite{Onishi2}. In agreement with the theory, the deformation parameter only slightly changes at changing the degree of viscoelasticity whereas the orientation angle is more sensitive.

\begin{figure}[tbp]
\includegraphics[scale=0.7]{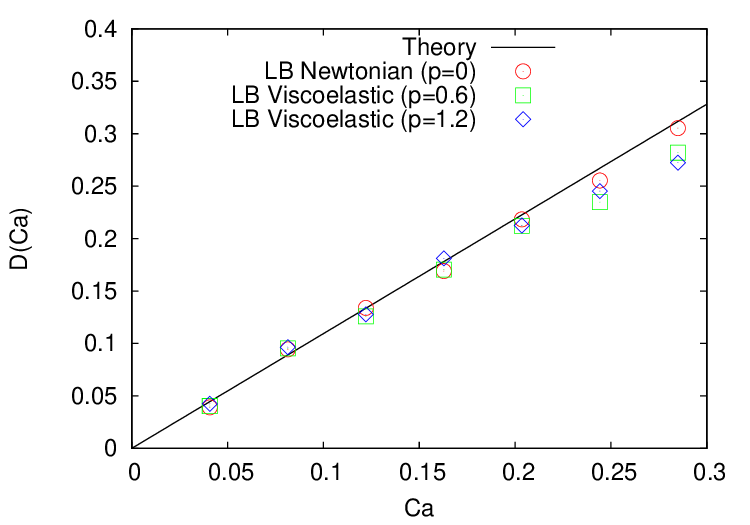}
\includegraphics[scale=0.7]{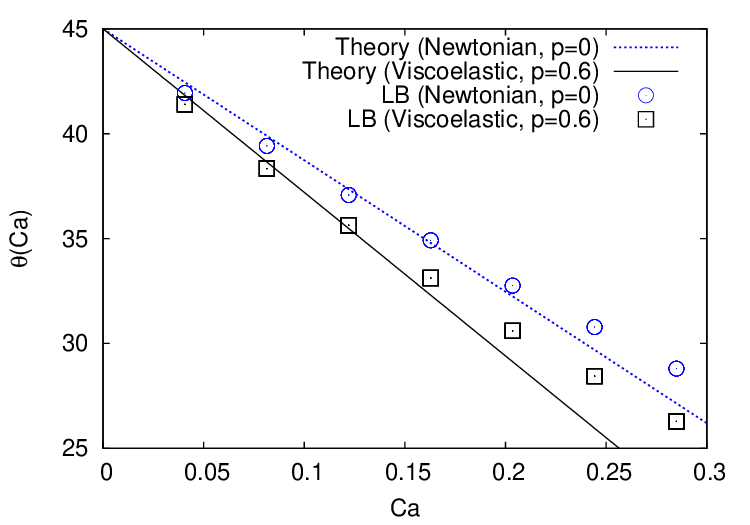}
\caption{Comparisons with the results of Onishi {\it et al.} \cite{Onishi2}. We plot the deformation parameter and the orientation angle obtained at steady state for different Capillary numbers $Ca$. The numerical parameters are chosen to be the same as those of Onishi {\it et al.} \cite{Onishi2}, see text for details. The solid lines and dashed lines are drawn with the theoretical predictions. To be noted that the Deborah number $De$ is denoted with $p$ in \cite{Onishi2}: we therefore decided to use $p$ to better (visually) establish a link with the results of \cite{Onishi2}.}
\label{fig:onishi2}
\end{figure}

Next, we compare the steady state shape of the drops in the different matrices in order to investigate viscoelasticity effects. Figures \ref{fig:onishi2} are the counterpart of figure 5 in \cite{Onishi2}: they report the steady state values of the deformation parameter and the orientation angle for different Capillary numbers. Note that the resolution used is already enough to achieve convergence to the hydrodynamic limit (see figure \ref{fig:resolution}). Indeed, in agreement with \cite{Onishi2}, the quantitative matching with the theoretical prediction by Greco \cite{Greco02} is achieved.

\section*{References}

\bibliography{archive}

\end{document}